\documentclass[preprint,aps,showpacs]{revtex4}
\usepackage[dvips]{graphicx}
\usepackage{bm}

\def\vk{\bm{k}}
\def\vkp{\bm{k}'}
\def\vp{\bm{p}}
\def\vpp{\bm{p}'}
\def\vq{\bm{q}}
\def\wpik{E_{\pi}(\bm{k})}
\def\enk{E_N(\bm{k})}
\def\edk{E_\Delta(\bm{k})}

\begin{document}



\title{Leading-order Corrections of a
 Dynamical Model for Electromagnetic Pion Production
Reactions}


\author{
T. Sato$^a$, 
T.-S. H. Lee$^b$
and T. Nakamura$^a$
}


\affiliation{
$^a$
Department of Physics, Osaka University, Toyonaka, Osaka 560-0043,
 Japan\\ 
$^b$
Physics Division, Argonne National Laboratory, Argonne,
Illinois 60439}


\begin{abstract}
By applying a unitary transformation method, we have
derived the leading-order corrections on the
effective Hamiltonian of a dynamical model developed in Phys. Rev. C{\bf 54},
2660 (1996) 
for electromagnetic pion production reactions.
The resulting 
$energy$-$independent$ one-loop corrections on the baryon masses 
and the $\gamma N \rightarrow \Delta$ vertex interaction are associated
 with the structure of the nucleon and $\Delta$ and have been calculated 
within a constituent quark model. We find that  
the one-loop  corrections on the magnetic M1 transition of the 
$\gamma N \rightarrow \Delta$ are very small, while
their contributions to the electric E2 and Coulomb C2
transitions are found to be
in opposite signs of that due to pion cloud effects
associated with the scattering states. 
Our results further indicate that the determination of 
the nonspherical
 $L=2$ components of the constituent
quark wavefunctions of $N$ and $\Delta$ from the
extracted empirical E2 and C2 form factors requires a rigorous and
complete calculation of meson cloud effects.
We also find that the one-loop
corrections on the non-resonant pion production operator
can resolve the difficulty in 
describing the near threshold $\gamma p \rightarrow \pi^0 p$ reaction.
Possible future developments are discussed. 
\end{abstract}


\maketitle

\section{Introduction}

In the past few years, extensive and precise data of electromagnetic meson
production reactions have become available and some of these data have been
used to extract the information about the nucleon
 resonances\cite{burkert}.
On the other hand, theoretical models for analyzing these reactions are still 
far from complete. Even in the simplest and well-studied $\Delta$ excitation
region, none of the most often applied models\cite{sl1,sl2,kamyan,dmt,maid}
 has been able to give $predictions$ which agree perfectly with
the single pion production data accumulated recently,
in particular the data on spin observables and
 longitudinal-transverse interference cross sections.
While these models can give an overall good description of fairly
extensive data, efforts must be made to remove the 
remaining discrepancies such that a complete
understanding of the $\Delta$ resonance can be obtained.
 The experiences gained from these efforts
will undoubtly be very useful for investigating the much more complex
higher mass $N^*$ resonances.
In this work, we report on the progress we have made in this direction,
focusing on the dynamical model we have developed
in Refs.\cite{sl1,sl2} (called the Sato-Lee (SL)  model
in the literatures).
In particular, we would like to explore 
how the bare $\gamma N \rightarrow \Delta$ parameters extracted
within the SL model
can be better understood in terms of the structure of $N$ and $\Delta$.
We would also like to see how the non-resonant pion production operator in the
SL model can be improved.

We first recall one of the most interesting results from the SL model. It was 
found that the pion cloud effects give very large contributions to
the $\gamma N \rightarrow \Delta$ transition form factors and is the source
of the differences between the values predicted by the
conventional constituent quark model and that extracted from
empirical amplitude analyses. The predicted very
pronounced $Q^2-$dependence in electric $E2$ and Coulomb $C2$ transitions have 
motivated several recent
experimental efforts. These pion cloud effects are calculated from
the following expression
\begin{eqnarray}
\bar{\Gamma}_{\gamma N, \Delta}(W, q) =\Gamma^0_{\gamma N,\Delta}(q)
+\int dk k^2  v^{tree}_{\gamma\pi}(q,k)
\frac{1}{W - E_\pi(k) - E_N(k) + i\epsilon}
 \bar{\Gamma}_{\pi N, \Delta}(W,k) \nonumber
\end{eqnarray}
where $\Gamma^0_{\gamma N,\Delta}$ is the bare vertex, 
$v^{tree}_{\gamma,\pi}$ is
the non-resonant $\gamma N \rightarrow \pi N$ amplitude calculated from
the standard Pseudo-Vector Born terms and the
$\rho$ and $\omega$ exchanges, and
$\bar{\Gamma}_{\pi N, \Delta}(W,k)$ is the dressed $\Delta \rightarrow \pi N$
vertex. One observes from the above equation
that these pion cloud effects are due to pions in the
$scattering$ states which can reach the on-shell momentum asymptotically.

We now examine how the above procedure is
related to our
current understanding of hadron structure. Because of the 
chiral symmetry
of QCD is spontaneously broken, it is generally believed that in the
region where the momentum transfer is not too large the structure of
the nucleon and $\Delta$ can be considered as systems made of constituent
quarks and virtual pions. We thus expect that their responses to the external
electromagnetic field can be from the constituent quarks and also from
the virtual pions. Obviously the pion-loop integration in the above
equation do not
account for all of the effects due to the virtual pions in hadrons. 
The leading term $\Gamma^0_{\gamma N, \Delta}$ must still contain some 
effects due to virtual pions which never go on-shell during the $N$-$\Delta$
transitions.
In this work, we will show how
the corrections due to these virtual pion cloud effects
can be derived by  applying the unitary transformation method. 
In a consistent derivation, the one-loop corrections on
the non-resonant pion production operator of the SL model have also been
derived. These one-loop corrections are also energy-independent and
are different from those due to pions in $scattering$ states.
These corrections are expected to have important effects in the
region where the pion electromagnetic reactions are sensitive to the
non-resonant amplitudes.

In section II, we recall 
a dynamical formulation within which the  leading order one-loop
corrections on the effective Hamiltonian of the SL model are derived. 
In section III, the consequences of  these leading order corrections on the
$\gamma N \rightarrow \Delta$ transitions are calculated and
interpreted within a constituent quark model. 
The  one-loop corrections on the non-resonant pion production 
operator are then investigated in section IV, focusing on
the s-wave amplitude of the near threshold $\pi^0$ photoproduction reaction.
Possible future developments are discussed in section V.

\section{Formulation}
As explained in Ref.\cite{sl1}, the SL model is constructed by 
applying  a 
 unitary transformation method to deduce from relativistic quantum field
theory an effective Hamiltonian for describing meson-baryon reactions.
The details of the employed unitary transformation 
has been given in Refs.\cite{sl1,ksh} and will not be repeated in this paper.
Here we only  emphasize that the starting point of the unitary
transformation method is a field theoretical Lagrangian density. This is
identical to other more familiar approaches for constructing dynamical
models of meson-baryon interactions, such as those based on
the ladder Bethe-Salpeter\cite{tjon,afnan}
or three-dimensional ladder Bethe-Salpeter
equations\cite{pearce,gross,hung,julich,pasca}. In the lowest order,
all approaches yield very similar, if not completely identical,
 scattering amplitudes.
Their differences are in the resulting
dynamical equations which are used to 
include nonperturbatively certain classes of higher order effects
that are deemed to be important for the processes considered.

\begin{figure}[h]
\centering
\includegraphics[width=2cm]{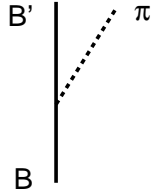}
\caption{The vertex interaction $\Gamma^0_{\pi B^\prime, B}$}
\label{fig1}
\end{figure}

\begin{figure}[h]
\centering
\includegraphics[width=6cm]{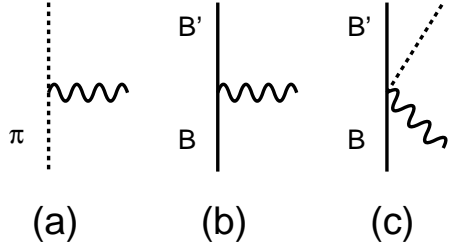}
 \caption{Electromagnetic interaction $H_{em}$ : (a) $J^\mu_\pi$,
(b) $J^\mu_{B^\prime, B}$, and (c) $J^\mu_{B^\prime, B,\pi}$.}
\label{fig2}
\end{figure}

To illustrate the unitary transformation method, it is sufficient
to  consider a model  Lagrangian density 
$L(\psi_N,\psi_\Delta,\phi_\pi)$
describing the pseudo-vector coupling between $\pi$, $N$ and
$\Delta$ fields. By using the standard canonical quantization procedure, a
Hamiltonian  can be constructed. 
To simplify the presentation, the spin and
isospin variables as well as the anti-particle components 
are suppressed here.
The resulting Hamiltonian can then be schematically
written as
\begin{eqnarray}
H=H_0 + H_I +H_{em} \, , \label{eqh1}
\end{eqnarray}
with
\begin{eqnarray}
H_0= \sum_{B}\int d\vp b^\dagger_{B}(\vp)b_B(\vp) E_B(\vp)
+\int d\vk a^\dagger_\pi(\vk)a_\pi(\vk) E_\pi(\vk) \nonumber
\end{eqnarray}
where $b^\dagger_B(\vp)(b_B(\vp))$ is the creation(annihilation)
operator for a baryon with momentum $\vp$, and
 $a^\dagger(\vk)(a(\vk))$  for
a pion with momentum $\vk$. The energy is defined as  
  $E_\alpha(\vp)=(m_\alpha+ \vp^2)^{1/2}$ with $m_\alpha$ denoting the
mass of particle $\alpha$.
Clearly, $H_0$ is the sum of free energy
operators for baryons($B= N, \Delta$)
and pion($\pi)$.
 The strong interaction Hamiltonian in Eq. (\ref{eqh1}) is
\begin{eqnarray}
H_I = \sum_{B,B'}[\Gamma^0_{\pi B', B}+ \mbox{h.c.}] \, , \label{eqh2}
\end{eqnarray}
with
\begin{eqnarray}
\Gamma^0_{\pi B',  B}=\int d\vp  d\vk
b^\dagger_{B'}(\vp - \vk)b_B(\vp)a^\dagger_\pi(\vk)
F_{\pi B',B}(\vp - \vk,\vk;\vp) , \label{eqh3}
\end{eqnarray}
where $F_{\pi B',B}(\vpp\vk;\vp)$ is
a vertex function describing the strength of the 
$\pi B \leftrightarrow B^\prime$
transition illustrated in Fig. \ref{fig1}. 
The corresponding electromagnetic interaction deduced from
 applying the minimum substitution on the considered
pseudo-vector coupling Lagrangian density  $L(\psi_N,\psi_\Delta,\phi_\pi)$
 can be written
as $H_{em}= \int d\bm{x} A\cdot J$, where $J^\mu$ is the current density 
operator and
 $A$ is the photon field.
The  resulting electromagnetic current can be schematically written as
\begin{eqnarray}
J^\mu = J^\mu_{\pi} + J^\mu_{B',B} + J^\mu_{B',B,\pi} , \label{eqh4}
\end{eqnarray}
where $J^\mu_{\pi}$, $J^\mu_{B',B}$, and $J^\mu_{B',B,\pi}$
define $\gamma \pi\rightarrow \pi$, $\gamma B \rightarrow B'$, and
the contact $\gamma B \rightarrow \pi B'$
transitions respectively, as illustrated in Fig. \ref{fig2}.
The details of these currents will be given in section III. 

The first step of the derivation is to
decompose the strong interaction Hamiltonian into two terms
\begin{eqnarray}
H_I=H_1^P + H_1^Q \, , \label{eqh5}
\end{eqnarray}
where
\begin{eqnarray}
H_1^P&=& \Gamma^0_{\pi N, \Delta} + \mbox{h.c.} \, ,\label{eqh6}\\
H_1^Q &=& [\Gamma^0_{\pi N ,N}
+\Gamma^0_{\pi \Delta,N}
+\Gamma^0_{\pi \Delta,\Delta}
] + \mbox{h.c.} \, . \label{eqh7}
\end{eqnarray}
Obviously, $H_1^P$ describes the physical process, while the processes
in $H_1^Q$ can not occur in free space because of the violation of
energy conservation. 
The second step is to perform unitary transformations on $H$ to construct
an effective Hamiltonian, which does not contain unphysical processes such
as those due to $H_1^Q$.
Keeping only the terms up the second order
in $H_I$, the resulting effective Hamiltonian is of the following form
\begin{eqnarray}
H_{eff}&  =& U^{\dagger}_2 U^{\dagger}_1 H U_1 U_2 \nonumber \\
       & = & H_0 + H_1^P + H_2^P + 
[ U^{\dagger}_2 U^{\dagger}_1 H_{em} U_1 U_2]
 + \Delta^Q \, , \label{eqh8}
\end{eqnarray}
where $U_n = \exp(i S_n)$ is the n-th unitary transformation 
with $S_n \propto (H_I)^n$, $H^P_1$
has been defined in Eq. (\ref{eqh6}) and
\begin{eqnarray}
 H^P_2= ([H_1^P,iS_1] + \frac{1}{2}[H_1^Q,iS_1])^P \label{eqh9}
\end{eqnarray}
Note that the commutators in Eq. (\ref{eqh9}) can generate both physical
and unphysical processes and only the terms for
 physical processes are kept in $H^P_2$.
The unphysical processes in $H_{eff}$ is contained in the last term of Eq. (\ref{eqh8})
\begin{eqnarray}
\Delta^Q = \{[H_0,iS_1] + H_1^Q \} + \{[H_0,iS_2] + H_2^Q \} \,, \label{eqh10}
\end{eqnarray} with
\begin{eqnarray}
H_2^Q =([H_1^P,iS_1] + \frac{1}{2}[H_1^Q,iS_1])^Q  \,. \label{eqh11}
\end{eqnarray}
$H^Q_2$ is defined by the same commutators in
$H^P_2$ except that only the unphysical processes
are kept here.

The desired effective Hamiltonian is obtained by eliminating the unphysical
processes $\Delta^Q$ Eq. (\ref{eqh8}). Obviously, this can be achieved by
imposing the following conditions
\begin{eqnarray}
{}[H_0,iS_1] + H_1^Q & = 0 \, ,\label{eqh12}\\
{}[H_0,iS_2] + H_2^Q & =0 \, . \label{eqh13}
\end{eqnarray}
To find $S_1$, consider the matrix elements of Eq. (\ref{eqh12}) 
between any two eigenstates
$\mid a>$ and $\mid b >$ of $H_0$; for example $H_0 \mid N> = E_N \mid N>$ and
$H_0\mid \pi N >= (E_\pi + E_N) \mid \pi N>$.
We then obtain a relation
 $(E_b - E_a) <a \mid i S_1 \mid b > = <a \mid H^Q_I \mid b >$, indicating
that  $S_1$ plays the same role as $H_I^Q$
in defining the interaction mechanisms.
It is then easy to verify that the general solution of Eq. (\ref{eqh12})
 can be written as the following operator form
\begin{eqnarray}
S_1 & = & -i \sum_{B',B} \int d\vp d\vk
    \frac{F_{\pi B',B}(\vp-\vk,\vk;\vp)}
         {E_B(\vp) - E_{B'}(\vp-\vk) - E_\pi(\vk)}
    \theta(m_\pi +m_{B'} - m_B)
    b^{\dagger}_{B'}(\vp-\vk) b_B(\vp) a^{\dagger}_\pi(\vk) \nonumber \\
   & + & \mbox{h.c.} \label{eqh14}
\end{eqnarray}
where the step function is defined as $\theta(x) =1(0)$, for $x >(<) 0$

\begin{figure}[h]
\centering
\includegraphics[width=5cm]{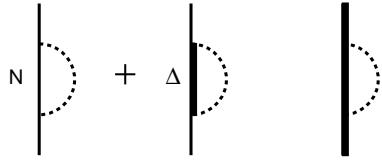}
 \caption{One-loop corrections $\Sigma^0_N$ and $\Sigma^0_\Delta$
on the nucleon and $\Delta$}
\label{fig3}
\end{figure}

For investigating $\pi N$ scattering and pion
photo- and electro-production at energies below two-pion production threshold,
it is sufficient to consider interactions defined within
 the Hilbert space $N\oplus\Delta\oplus
\pi N\oplus \gamma N$.
By using  Eq. (\ref{eqh3}) and Eq. (\ref{eqh14}), we
can evaluate  the matrix elements of $H^P_2$, defined by Eq. (\ref{eqh9}), 
between two one-baryon states. This will
generate the one-loop corrections, $\Sigma^0_N$ and
$\Sigma^0_\Delta$, to the masses of $N$ and
$\Delta$, as illustrated in Fig. \ref{fig3}. Explicitly, we find
\begin{eqnarray}
\Sigma_N^0 &=& \frac{i}{2}\sum_{ B^\prime= N,\Delta}
[< N\mid  \Gamma^{0\dagger}_{\pi B',N}\mid \pi B^\prime>
<\pi B^\prime \mid S_1 \mid N> \nonumber \\
&-& < N\mid  S_1 \mid \pi B^\prime>
<\pi B^\prime \mid\Gamma^0_{\pi B',B} \mid N>],  \label{eqh15}\\
\Sigma_\Delta^0 &=& \frac{i}{2}
[< \Delta\mid  \Gamma^{0\dagger}_{\pi \Delta,\Delta}\mid \pi \Delta>
 <\pi \Delta \mid S_1 \mid \Delta> \nonumber \\
&-& < \Delta\mid  S_1 \mid \pi \Delta>
 <\pi\Delta \mid \Gamma^0_{\pi \Delta,\Delta} \mid \Delta>]. \label{eqh16}
\end{eqnarray}
Using the solution Eq. (\ref{eqh14}) for $S_1$ to evaluate
 the above two equations, we will get expressions
involving one-loop integrations over $energy-independent$ propagators which
are also specified in Eq. (\ref{eqh14}). The detailed forms
 will be given in the next section where
we will perform calculations using a model for the vertex interaction 
$\Gamma^0_{\pi B',B}$.
Note that $\Sigma^0_\Delta$ does not include the loop over intermediate
$\pi N$ state since the effects due to
$\Delta \rightarrow \pi N$ is already accounted for
by $H^P_1$ of Eq. (\ref{eqh6}) and must be excluded in $Q$ interactions.

\begin{figure}[h]
\centering
\includegraphics[width=5cm]{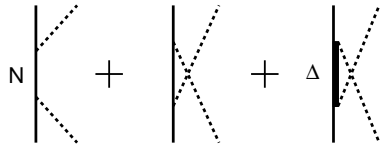}
 \caption{$\pi N$ interactions.}
\label{fig4}
\end{figure}

Taking the expectation value of $H^P_2$ between two $\pi N$ states,
we then generate the $\pi N$ potential $v_{\pi N}$,
illustrated in in Fig. \ref{fig4}. Extending the procedure described
above to also 
include spin and isospin indices as well as the anti-particle components
and $\rho$ meson, the matrix elements of 
$v_{\pi N}$ given explicitly in the SL model can  then be
obtained. On the other 
hand, the one-loop corrections $\Sigma^0_N$ and
$\Sigma^0_\Delta$ are not treated explicitly in SL model.
  
\begin{figure}[h]
\centering
 \includegraphics[width=2cm]{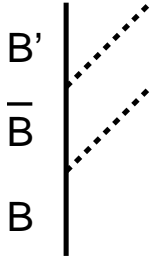}
 \caption{Unphysical processes due to $H_2^Q$.}
 \label{fig5}
\end{figure}

To determine $S_2$ from Eq. (\ref{eqh13}), we need to know the mechanisms
contained in $H_2^Q$ defined by Eq. (\ref{eqh11}). 
With the solution Eq. (\ref{eqh14}) for $S_1$, one can easily see that
$H_2^Q$ can generate the unphysical $B \rightarrow \pi\pi B$ processes illustrated
in Fig. \ref{fig5}.
With the similar procedure employed in solving Eq. (\ref{eqh12}) for $S_1$,
we find that the solution of Eq. (\ref{eqh13}) from eliminating the unphysical processes
illustrated in Fig. \ref{fig5} can be written explicitly as the following
operator form
\begin{eqnarray}
S_2 & = &-i \sum_{B',B,\bar{B}} \int d\vp d\vk d\vkp
\frac{F_{\pi B^\prime,\bar{B}}(\vp-\vk-\vkp,\vkp;\vp-\vk)
    F_{\pi \bar{B},B}(\vp-\vk,\vk;\vp) }
     {E_B(\vp) - E_{B'}(\vp-\vk-\vkp) - E_\pi(\vk)-E_\pi(\vkp)}
    \theta(2 m_\pi + m_{B'} - m_B)
  \nonumber \\
& &  {}  [
\frac{\theta(m_\pi + m_{B'} - m_{\bar{B}})}
     {E_{B'}(\vp-\vk-\vkp) - E_{\bar{B}}(\vp-\vk) + E_\pi(\vkp)}
(\frac{\theta(m_\pi + m_{\bar{B}}-m_B)}{2} + \theta(-m_\pi - m_{\bar{B}} + m_B))
 \nonumber \\
& &
 + \frac{\theta(m_\pi + m_{\bar{B}} - m_{B})}
     {E_B(\vp) - E_{\bar{B}}(\vp-\vk) - E_\pi(\vk)}
(\frac{\theta(m_\pi + m_{B'} - m_{\bar{B}})}{2}
 + \theta(-m_\pi - m_{B'} + m_{\bar{B}}))
 ]
 \nonumber \\
 & &
 b^{\dagger}_{B'}(\vp-\vk-\vkp) b_B(\vp)
 a^{\dagger}_\pi(\vkp) a^{\dagger}_\pi(\vk) + \mbox{h.c.} \label{eqh17}
\end{eqnarray}
We note that $S_2$ does not play any role in generating effective
the Hamiltonian  up to the second order in
$\Gamma^0_{\pi B, B'}$. But it is needed to evaluate the effective 
electromagnetic interaction 
operator defined by 
the term $[ U^{\dagger}_2 U^{\dagger}_1 H_{em} U_1 U_2] $
in Eq. (\ref{eqh8}).

\begin{figure}[h]
\centering
\includegraphics[width=10cm]{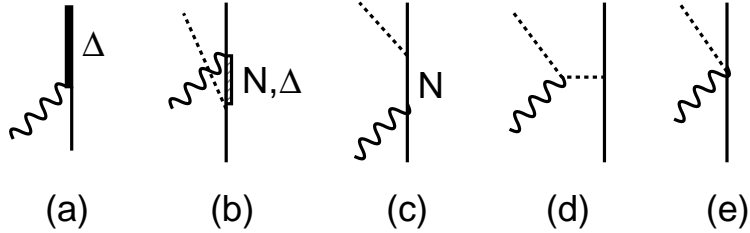}
 \caption{Leading order terms of pion photoproduction:
$\Gamma^0_{\gamma N,\Delta}$=(a), $v^{tree}_{\gamma \pi} = (b)+(c)+(d)+(e)$
}
\label{fig6}
\end{figure}

Keeping only the terms up to the same second order in $H_I$,  
we can write $[ U^{\dagger}_2 U^{\dagger}_1 H_{em} U_1 U_2] =
 \int d\bm{x} A\cdot J_{eff}$ with the
effective current defined by
\begin{eqnarray}
J^\mu_{eff} & = &
         J^\mu       + [J^\mu,iS_1]      + \frac{1}{2}[[J^\mu,iS_1],iS_1]
      + [J^\mu,iS_2] + [[J^\mu,iS_1],iS_2] \,.
\label{eqh18}
\end{eqnarray}
By using the properties of $S_1$ and the electromagnetic
coupling illustrated in Fig. \ref{fig2}, one can see that
the first two terms of Eq. (\ref{eqh18}) for $J^\mu_{eff}$
generate the tree mechanisms shown in Fig. \ref{fig6}. Explicitly, we can
see the following correspondences :
\begin{eqnarray}
\mbox{Fig. 6a} &:& < \Delta \mid J^\mu_{\Delta,N}\mid \gamma N >,\label{eqh19}\\
(\mbox{Fig.6b + Fig.6c})&:&<\pi N \mid [J^\mu_{N,N}, iS_1]
            + J^\mu_{N,\Delta} iS_1 \mid \gamma N >,\label{eqh20}
\\
\mbox{Fig.6d} &:&<\pi N \mid [J^\mu_{\pi}, iS_1]\mid \gamma N>,  \label{eqh21}\\
\mbox{Fig.6e} &:&<\pi N \mid J^\mu_{N,N,\pi} \mid \gamma N> .\label{eqh22}
\end{eqnarray}
Extending the procedure described above to also
include spin and isospin indices as well as the anti-particle components
and $\rho$ and $\omega$ meson-exchange, the matrix elements for
$v_{\gamma N}$ given explicitly in the SL model can  then be obtained.

\begin{figure}[h]
\centering
\includegraphics[width=11cm]{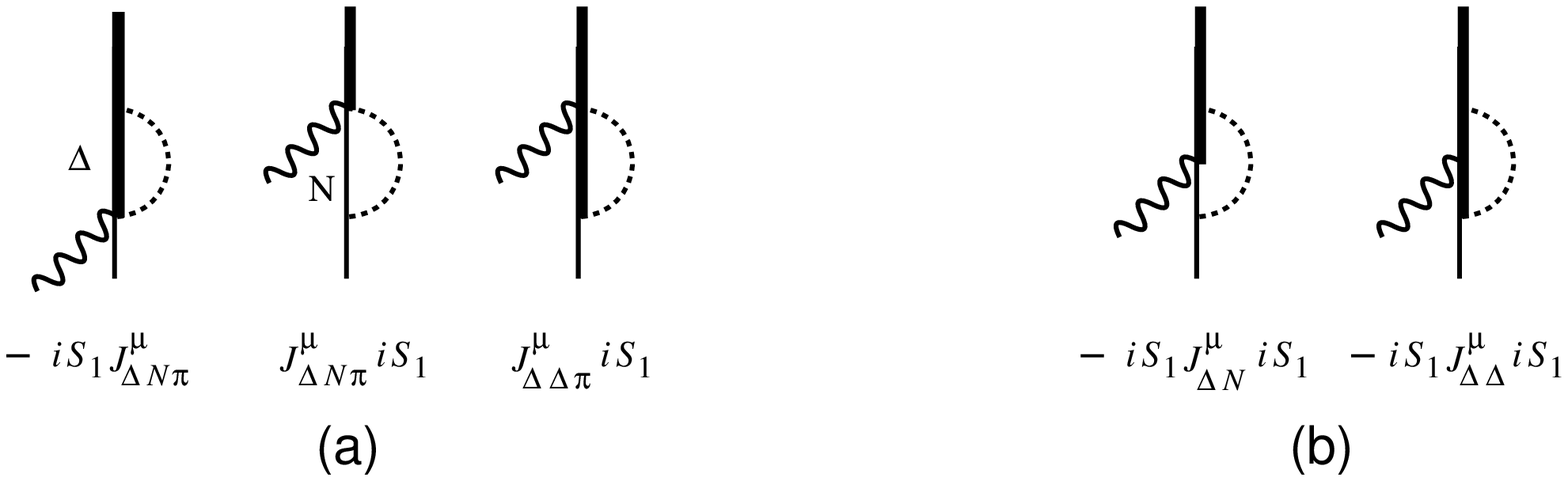}

\vspace*{0.5cm}
\centering
\includegraphics[width=13cm]{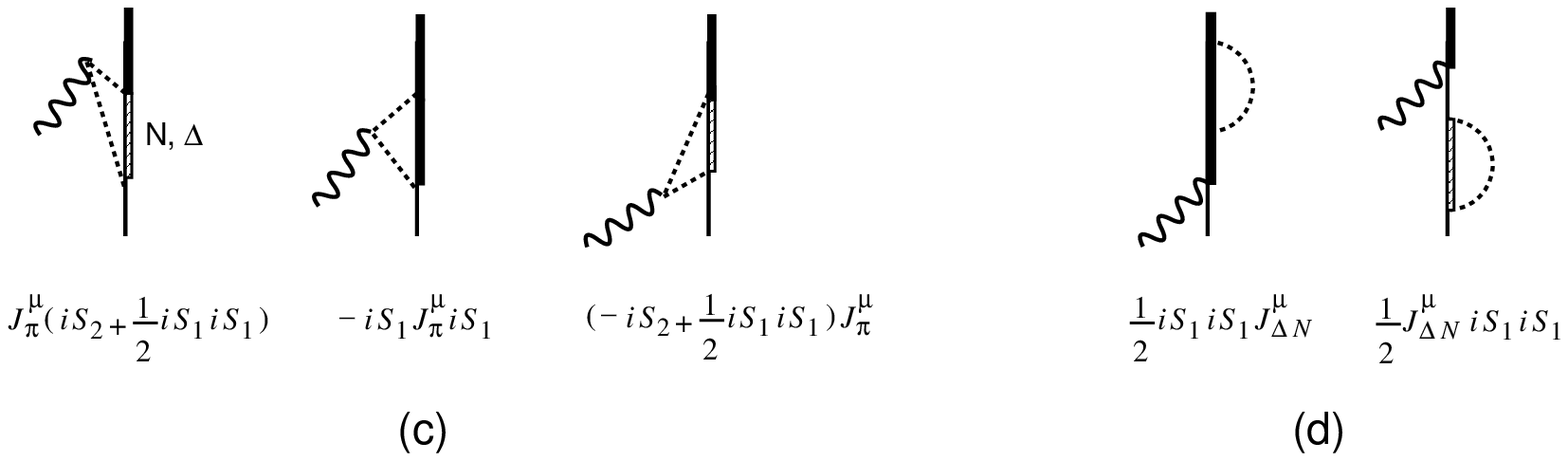}
\caption{Loop Correction $\Gamma^{1-loop}_{\gamma N,\Delta}$
 on $\gamma N \rightarrow \Delta$
transition.}
\label{fig7}
\end{figure}

The one-loop corrections
on the $\gamma B \rightarrow B^\prime$ vertex and 
non-resonant $\gamma N \rightarrow \pi N$ amplitude can be generated from
the following operators in Eq. (\ref{eqh18}), 
\begin{eqnarray}
J^{\mu,1-loop}= 
        [J^\mu,iS_1]      + \frac{1}{2}[[J^\mu,iS_1],iS_1]
      + [J^\mu,iS_2] + [[J^\mu,iS_1],iS_2] \,.
\label{eqh23}
\end{eqnarray}
For $\gamma N \rightarrow \Delta$, the possible intermediate states involved
in evaluating the one-loop corrections
are illustrated in Fig. \ref{fig7} with the following correspondences:
\begin{eqnarray}
\mbox{Fig.7a} &:& <\Delta \mid  [ J^\mu_{B',B, \pi}, iS_1]\mid \gamma N>,
  \label{eqh24} \\
\mbox{Fig.7b} &:& <\Delta \mid - iS_1J^\mu_{B',B}iS_1\mid \gamma N>, 
\label{eqh25}\\
\mbox{Fig.7c} &:& <\Delta \mid
\frac{1}{2}[[J^\mu_{\pi},iS_1],iS_1] + [J^\mu_\pi, iS_2] \mid \gamma N>, 
\label{eqh26}\\
\mbox{Fig.7d} &:&<\Delta \mid \frac{1}{2}(J^\mu_{B',B} iS_1 iS_1
 +  iS_1 iS_1 J^\mu_{B',B} ) \mid \gamma N >.
\label{eqh27}
\end{eqnarray}
Similar  expressions and diagrams are also for the one-loop corrections,
$< N \mid J^{1-loop}_\mu \mid \gamma N >$, for the nucleon electromagnetic form 
factors. 

The one-loop corrections on the non-resonant $\gamma N \rightarrow \pi N$ 
amplitudes can also be obtained by  taking the matrix element of
$J_\mu^{1-loop}$ between $\pi N$ and $\gamma N$ states. 
We will elaborate this more complex object in section IV.

With the above derivations, the effective Hamiltonian Eq. (\ref{eqh8}) within
the subspace $N\oplus\Delta\oplus \pi N \oplus \gamma N$
can be written as
\begin{eqnarray}
H_{eff} = [H_0 + \Sigma^0_N + \Sigma^0_\Delta]
+ [\Gamma_{\pi N, \Delta} + \Gamma_{\gamma N, \Delta}]
  +[v_{\pi N}+ v_{\gamma \pi}]. \label{eqh28}
\end{eqnarray}
The mass correction terms, $\Sigma^0_N$ and $\Sigma^0_\Delta$, are
illustrated in Fig. \ref{fig3}.
The vertex interactions in Eq. (\ref{eqh28}) are
\begin{eqnarray}
\Gamma_{\pi N, \Delta} &=& \Gamma^0_{\pi N, \Delta},\label{eqh29} \\
\Gamma_{\gamma N, \Delta}
 &=& \Gamma^0_{\gamma N,\Delta}
+ \Gamma^{1-loop}_{\gamma N,\Delta}. \label{eqh30}
\end{eqnarray}
Here we have defined
\begin{eqnarray}
\Gamma^0_{\gamma N, \Delta} = \int  d\bm{x} <\Delta \mid
 A \cdot   J^\mu_{\Delta, N} \mid \gamma N >. \label{eqh31}
\end{eqnarray}
The one-loop corrections
$\Gamma^{1-loop}_{\gamma N, \Delta}$
are  defined by Eqs. (\ref{eqh24})-(\ref{eqh27}) and illustrated in Fig. \ref{fig7}.
Note that up to the second order in $H_I$,
there is no one-loop correction to the $\pi N \rightarrow \Delta$ vertex
in Eq. (\ref{eqh29}). 

The $\pi N$ potential $v_{\pi N}$ in Eq. (\ref{eqh28})
 is illustrated  in Fig. \ref{fig4}. The non-resonant $\gamma N \rightarrow \pi N$
transition interaction is defined by
\begin{eqnarray}
v^{\gamma \pi} = v^{tree}_{\gamma \pi} + v^{1-loop}_{\gamma \pi}  \label{eqh32}
\end{eqnarray}
where 
$v^{tree}_{\gamma \pi}$ is defined by Eqs. (\ref{eqh19})-(\ref{eqh22}) and illustrated in
Figs. (6b)-(6d), and $v^{1-loop}_{\gamma \pi}$ is the one-loop corrections 
which can be calculated by taking the matrix element of Eq. (\ref{eqh23}) 
between $\gamma N$ and $\pi N$ states.

The SL model can be obtained from the
effective Hamiltonian $H_{eff}$ of Eq. (\ref{eqh28}) by making the following
simplifications.
First, the mass correction terms $\Sigma^0_N$ and $\Sigma^0_\Delta$
are not treated explicitly and are included in the
physical nucleon mass $m_N$=938.5 MeV and $m_\Delta =1299$ MeV  determined
in Ref.\cite{sl1}. 
Second, the one-loop corrections $\Gamma^{1-loop}_{\gamma N, \Delta}$ are not
calculated explicitly and $\Gamma_{\gamma N,\Delta}$, instead of
$\Gamma^0_{\gamma N,\Delta}$, is adjusted to fit the data.
Finally, the non-resonant   
$v^{1-loop}_{\gamma\pi}$ is neglected. 

The above derivation indicates that the method of unitary transformation
has provided a systematic way to improve the SL model. 
In the next two sections,we will explore the consequences 
of these leading-order corrections derived in this section.

\section{One-loop corrections on the one-baryon processes}

To evaluate
the one-loop corrections Eqs. (\ref{eqh15})-(\ref{eqh16}) for the baryon masses
and Eqs. (\ref{eqh24})-(\ref{eqh27}) for the $\gamma N \rightarrow N$ and
$\gamma N \rightarrow \Delta$ transitions, we need to define the vertex
function $F_{\pi B',B}$ of Eq. (\ref{eqh3}) and the matrix elements of currents of
Eq. (\ref{eqh4}). As an exploratory step, we assume that these can be calculated
from a  model within which the
pion is coupled to constituent quarks by the usual
pseudo-vector coupling and the electromagnetic interaction is
introduced by the minimum substitution. We further assume that
the constituent quarks in $N$ and $\Delta$ are nonrelativistic and
only have $L=0$ s-wave configurations. Accordingly,
the usual nonrelativistic limit is also taken to define the couplings of
$\pi$ and $\gamma$ with constituent quarks.
With these simplifications, 
we can cast the resulting $\pi B \rightarrow B'$ vertex into the following
form 
\begin{eqnarray}
F_{\pi^i B,B'}(\vp, \vk;\vpp)
=\frac{i}{\sqrt{(2\pi)^3}}\frac{1}{\sqrt{2E_\pi (k)}}
\frac{f_{\pi BB'}}{m_\pi}
            (\bm{S}_{B',B}\cdot \vk)
            (\bm{T}_{B',B}\cdot \bm{I}^i_\pi ) F_{B',B}(\bm{k}) \label{eqh33}
\end{eqnarray}
Here  $\bm{I}^i_\pi$ is a vector
associated with the pion isospin state $i$, and $F_{B',B}(\bm{k})$
is a form factor calculated from quark wave functions.
The spin and isospin operators $\bm{S}_{B'B},\bm{T}_{B'B}$
are  defined as follows. For diagonal spin operators they are
twice of the spin angular momentum
operator.
\begin{eqnarray}
\bm{S}_{NN} & = & 2\bm{J} = \bm{\sigma}, \label{eqh34}\\
\bm{S}_{\Delta\Delta} & = & 2\bm{J} = \bm{S}_{\Delta}; \label{eqh35}
\end{eqnarray}
while the transition spin operators are defined as
\begin{eqnarray}
\bm{S}_{\Delta N} & = & \bm{S}, \label{eqh36}\\
\bm{S}_{N \Delta} & = & \bm{S}^{\dagger}. \label{eqh37}
\end{eqnarray}
Within the considered
SU(6) quark model, these operators are related to each other 
, as given explicitly in Table I.
\begin{table}[htb]
\caption[]{Coupling constants. Here
 $\mu_N^S = \mu_P/6,\mu_N^V=5\mu_P/6$ with
$\mu_P=e/2m_q$. $m_q= 360$ MeV is the quark mass which is determined
 here by including
the one-loop corrections to fit the proton magnetic moment (see Table II).}
\begin{tabular}{cccccc}
 $B'$ $B$ & $\bm{S}_{B'B}$($\bm{T}_{B'B}$)
             & $<B'||\bm{S}_{B'B}||B>$ & $f_{\pi B'B}$
              & $\mu_B^S$ &  $\mu_{B'B}^V$ \\ \hline
 $NN$ & $\bm{\sigma}$ & $\sqrt{6}$ &  $f_{\pi NN}$
          & $\mu_N^S$ & $\mu_N^V$ \\ 
 $\Delta N$ & $\bm{S}$ & $2$ & $\sqrt{72/25}f_{\pi NN}$
          & $0$       & $\sqrt{72/25}\mu_N^V$\\
 $N \Delta $ & $\bm{S}^{\dagger}$ & $-2$ & $\sqrt{72/25}f_{\pi NN}$
          & $0$       & $\sqrt{72/25}\mu_N^V$\\
 $\Delta \Delta$ & $\bm{S}_{\Delta}$ & $2\sqrt{15}$ & $f_{\pi NN}/5$
          & $\mu_N^S$   & $\mu_N^V/5$ \\ \hline
\end{tabular}
\end{table}
The same table also define the reduced matrix elements for
isospin operators $\bm{\tau}$ for $NN$,
$\bm{T}$ for $\Delta N$, $\bm{T}^\dagger$ for $N\Delta$, and $\bm{T}_\Delta$ for 
$\Delta\Delta$.
                                                                                
 Let us first calculate the mass correction terms 
$\Sigma^0_N$ and $\Sigma^0_\Delta$ that are given in Eqs. (\ref{eqh15})-(\ref{eqh16}).
By using Eqs. (\ref{eqh3}), (\ref{eqh14}) and (\ref{eqh33}),
 we obtain in the rest frame of $N$ and
$\Delta$
\begin{eqnarray}
\Sigma^0_N & = &  
\int \frac{d\bm{k}}{(2\pi)^3} <m_{s_N} m_{\tau_N}\mid[
  (\frac{f_{\pi NN}}{m_\pi})^2 \frac{1}{2\wpik}
    \frac{\bm{\sigma}\cdot\bm{k}\bm{\sigma}\cdot\bm{k}
      \bm{\tau}  \cdot\bm{\tau} \mid F_{N,N}(k)\mid^2} {m_N - \wpik - \enk}
 \nonumber \\
 & +  & (\frac{f_{\pi N\Delta}}{m_\pi})^2\frac{1}{2\wpik}
\frac{\bm{S}\cdot\bm{k}\bm{S}^{\dagger}\cdot\bm{k}
      \bm{T}\cdot\bm{T}^{\dagger} \mid F_{N,\Delta}(k)\mid^2}
{m_N - \wpik - \edk}]
\mid m_{s_N} m_{\tau_N} >, \label{eqh38}\\
\Sigma^0_\Delta & = &
\int \frac{d\bm{k}}{(2\pi)^3}
<m_{s_\Delta} m_{\tau_\Delta}\mid
    (\frac{f_{\pi \Delta\Delta}}{m_\pi})^2 \frac{1}{2\wpik}
  \frac{\bm{S}_{\Delta}\cdot\bm{k}\bm{S}_{\Delta}\cdot\bm{k}
         \bm{T}_{\Delta}\cdot\bm{T}_{\Delta}
\mid F_{\Delta,\Delta}(k)\mid ^2}
{m_\Delta - \wpik - \edk}
\mid m_{s_\Delta} m_{\tau_\Delta} >. \nonumber \\
& & \label{eqh39}
\end{eqnarray}

To perform the calculations, we need to define the form factor
$F_{B,B^\prime}(\bm{k})$ in Eq. (\ref{eqh33}).
To be consistent with  the SL model, we here depart from the usual
oscillator form and 
set $F_{B,B'}(\bm{k}) = (\Lambda^2/(\Lambda^2+\vk^2))^2$ with $\Lambda=650$
(MeV/c)$^2$ for all $\pi BB^\prime$ vertices.
Eqs. (\ref{eqh38})-(\ref{eqh39}) then lead to the following results
\begin{eqnarray}
\Sigma^0_N &=& \Sigma^0_N(\pi N) + \Sigma^0_N(\pi\Delta) \nonumber \\
         &=& -73.5 MeV -65.4MeV = -139. MeV \nonumber \\
\Sigma^0_\Delta&=& \Sigma^0_\Delta(\pi\Delta) \nonumber \\
             &=&-76.6 MeV  \label{eqh40}
\end{eqnarray}
Here we indicate the intermediate state in each pion loop term.
Note that $\pi N$ intermediate state is excluded in the
correction to the bare $\Delta$ mass, since its effect is already included
in  the rescattering term induced by the vertex interaction
$\Gamma_{\pi N,\Delta}$ of Eq. (\ref{eqh29}).
The contribution of this rescattering
to the mass shift is 
\begin{eqnarray}
\Sigma^{res}_\Delta & = &
P \int \frac{d\bm{k}}{(2\pi)^3}
<m_{s_\Delta} m_{\tau_\Delta}\mid
    (\frac{f_{\pi N\Delta}}{m_\pi})^2 \frac{1}{2\wpik}
  \frac{\bm{S}^\dagger\cdot\bm{k}\bm{S}\cdot\bm{k}
         \bm{T}^\dagger\cdot\bm{T}
\mid F_{N,\Delta}(k)\mid ^2}
{m_\Delta - \wpik - \enk}
\mid m_{s_\Delta} m_{\tau_\Delta} >  \nonumber \\
&=& -46.2 MeV \label{eqh41}
\end{eqnarray}
where $P$ denotes taking the principal-value part of the integration.
                                                                                
We now note that $(H_0 + \Sigma^0_N + \Sigma^0_\Delta)$ of the effective
Hamiltonian Eq. (\ref{eqh28}) defines the physical nucleon mass and
the pole position $W_\Delta = m_\Delta =1232$ MeV
 of the K-matrix of $\pi N$ scattering in
$P_{33}$ channel.
Thus, we have the following relations 
\begin{eqnarray}
m_N = m_N^0 + \Sigma_N^0 \,, \label{eqh42}
\end{eqnarray}
and 
\begin{eqnarray}
W_\Delta = m_\Delta^0 + \Sigma_\Delta^0  
+ \Sigma_\Delta^{res} \,. \label{eqh43}
\end{eqnarray}

With the above results, the mass parameters $m_N^0$
and $m_\Delta^0$ associated with $H_0$
of the effective Hamiltonian Eq. (\ref{eqh28}) can then
be determined  
\begin{eqnarray}
m_N^0 & = & m_N - \Sigma_N^0 = 1077MeV \nonumber \\
m_\Delta^0 & = & W_\Delta - \Sigma_\Delta^0 - \Sigma_\Delta^{Res}=1355MeV
 \nonumber
\end{eqnarray}
The difference of the two bare masses are
\begin{eqnarray}
\delta^0 = m_\Delta^0 - m_N^0 = 278MeV \label{eqh44}
\end{eqnarray}
The mass parameters $m^0_N$ and $m^0_\Delta$ obtained above can be
considered as data for determining the parameters of
a hadron structure model which `exclude' the  pion degree
of freedom. Accordingly, one can assume that  $H_0$ of
Eq. (\ref{eqh28}), which is defined by these two bare masses, can be
identified with a model Hamiltonian defining the structure of the
constituent quarks within the nucleon and $\Delta$.

Most of the existing  constituent quark model
calculations\cite{isgu,caps,iach,gloz},
determine their parameters by fitting the mass difference
$\delta m = m_\Delta(=1232) - m_N(=938.5) = 294$ MeV, not by
reproducing the absolute values of the masses of $N$ and $\Delta$.
We note that this mass difference is not so different from that given in 
Eq. (\ref{eqh44}). Thus we can identify $H_0$ of Eq. (\ref{eqh28}) as
the  constituent quark model Hamiltonian with its eigenfunctions
$\mid N >$ and $\mid \Delta >$
consisting  of three quarks. Accordingly, the current matrix element
$\Gamma^0_{\gamma N,\Delta}$ defined by Eq. (\ref{eqh31}) can be identified with
the prediction from the constituent quark models.

We now turn to calculating the electromagnetic form factors. The constituent
quark
contribution is described by the current operator $J_{B,B^\prime}$ of
Eq. (\ref{eqh4}). Within the considered SU(6) constituent quark model  
and in the second quantization notation of Eq. (\ref{eqh3}), we can write
\begin{eqnarray}
 \bm{J}_{B',B}  & = & \int \frac{d\vp d\vpp d\vq}{(2\pi)^3} [
  \delta(\vp - \vq - \vpp) e^{i\vq\cdot \bm{x}}
 b^\dagger_{B^\prime}(\vpp) b_B(\vp) 
\nonumber \\
&\times& (e [\frac{1}{2} + \frac{T_{B',B}^z}{2}]
  \frac{\bm{p} + \bm{p}'}{2m_{B}}\delta_{B',B}
  + [\mu_{B}^S \delta_{B',B}
   + \mu_{B',B}^V T_{B',B}^Z] i \bm{S}_{B',B} \times \bm{q})
    F^{em}_{B^\prime,B}(q^2)
\nonumber \\
& & + (\mbox{h.c.})]  \label{eqh45}
\end{eqnarray}
where $F^{em}_{B,B^\prime}(q^2)$ is an electromagnetic
form factor, and the parameters
$\mu^S_B$ and $\mu^V_{B^\prime B}$ are defined in Table I in terms of
  $\mu_P = {e}/(2m_q)$ with $m_q$ denoting the quark
mass. In consistent with the SL
 model, the other two current operators in Eq. (\ref{eqh4}) are
\begin{eqnarray}
 \bm{J}_{\pi} & = &
\sum_{i,j} \int \frac{d\vk d\vkp d\vq}{(2\pi)^3} [
  \delta(\vk - \vq - \vkp) e^{i\vq\cdot \bm{x}}
\frac{1}{\sqrt{2E_\pi(\vk)}}\frac{1}{\sqrt{2E_\pi(\vkp)}}
   a^\dagger_{\pi^i}(\vkp) a_{\pi^j}(\vk)
\nonumber \\
&\times& 
( -i e \epsilon_{ij3}(\vk+\vkp)) F^{em}_{\pi}(q^2) + (\mbox{h.c.})] \label{eqh46}
\end{eqnarray}
and
\begin{eqnarray}
 \bm{J}_{B',B,\pi} & = &
\sum_{i,j} \int \frac{d\vp d\vpp d\vk d\vq} {(2\pi)^{9/2}}
  \delta(\vp - \vq - \vpp - \vk) e^{i\vq\cdot \bm{x}}
\frac{1}{\sqrt{2E_\pi(\vk)}}
b^\dagger_{B^\prime}(\vpp)a^\dagger_{\pi^i}(\vk) b_B(\vp) 
\nonumber \\
&\times &
( e \frac{f_{\pi B',B}}{m_\pi} \epsilon_{ij3}T_{B',B}^j
  \bm{S}_{B',B})F^{em}_{B^\prime,B\pi}(q^2)
 + (\mbox{h.c}) ]. \label{eqh47}
 \end{eqnarray}
The corresponding charge density operators are
\begin{eqnarray}
 \rho_{\pi} & = & 
    \sum_{i,j} \int \frac{d\vk d\vkp d\vq} {(2\pi)^3}
 \frac{1}{\sqrt{2E_\pi(\vk)}}\frac{1}{\sqrt{2E_\pi(\vkp)}}F^{em}_{\pi}(q^2)
    e^{i\vq\cdot \bm{x}}
   \nonumber \\
 & & {}[
  \delta(\vk - \vq - \vkp) 
   a^\dagger_{\pi^i}(\vkp) a_{\pi^j}(\vk)
   (-ie \epsilon_{ij3}(E_\pi(\bm{k})+E_\pi(\bm{k}^\prime))) \nonumber \\
  & + &  
  \delta(\vk - \vq + \vkp) 
   a_{\pi^i}(\vkp) a_{\pi^j}(\vk)
   (-ie \epsilon_{ij3}(E_\pi(\bm{k})-E_\pi(\bm{k}^\prime))) 
 +(\mbox{h.c.})], \label{eqh48}\\
  \rho_{B^\prime,B}   & = &
   \delta_{B^\prime,B} \int  \frac{d\vp d\vpp d\vq}{(2\pi)^3} [
  \delta(\vp - \vq - \vpp)  e^{i\vq\cdot \bm{x}}
   b^\dagger_{B^\prime}(\vpp) b_{B}(\vp)
   e\frac{1+T^z_{B',B}}{2}F^{em}_{B^\prime,B}(q^2)+(\mbox{h.c.})] \label{eqh49}
\end{eqnarray}
In the above equations, the form factors $F^{em}_{B^\prime,B}(q^2)$, 
$F^{em}_{\pi}(q^2)$
and $F^{em}_{B^\prime,B\pi}(q^2)$ should in principle be calculated from the
associated hadron structure. This is a nontrivial task, as well recognized.
For this exploratory study, we simply set all of these form factors equal to
$F_V(q^2) = 1/(1-q^2/M_V^2)^2$ with $M_V = 0.76$ GeV being the mass of vector
meson. Obviously, this is the simplest prescription to maintain the gauge 
invariance. 

With the above definitions, we can evaluate loop corrections
defined by Eqs. (\ref{eqh24})-(\ref{eqh27}) by inserting appropriate intermediate
states(illustrated in Fig. 7) and using  Eq. (\ref{eqh14}) for
$S_1$ and Eq. (\ref{eqh17}) for $S_2$.
 
To proceed,  we need to first fix the quark mass $m_q$
which determine the current
$J_{B^\prime B}$.
This is done by fitting the nucleon magnetic moments.
The one-loop corrections (similar to what are shown in Fig. \ref{fig7}) are included
in the fit.
We find that the nucleon magnetic moments can be
reproduced very well if we set the quark mass as $m_q=360$
MeV. The results for the magnetic moments are shown in
Table II.
We see that the loop corrections are about 5$\%$ for proton and
10 $\%$ for neutron.
\begin{table}
\caption[]{Magnetic moment of nucleon in unit $\mu_N$}
\begin{tabular}{ccc}
        &   'tree' & \hspace*{0.5cm}Total with loop corrections \\ \hline
proton  &   2.61   &  2.75 \\
neutron &  -1.74   & -1.95
\end{tabular}
\end{table}
It is important to note that the size of one-loop corrections depend heavily on
the range $\Lambda$ of the form factor $F_{B^\prime,B}$ of Eq. (\ref{eqh33}).

We now turn to investigating the loop corrections on
the $\gamma N \rightarrow \Delta$ transition.
Following the formulation presented in SL model,
the $\gamma N \rightarrow \Delta$ vertex function calculated in the $\Delta$
rest frame can be written in the following form
\begin{eqnarray}
<\Delta\mid \Gamma_{\gamma N \rightarrow \Delta} \mid q>
  &=& -\frac{e}{(2\pi)^{3/2}}\sqrt{\frac{E_N(\bm{q})+m_N
}{2E_N(\bm{q})}}
       \frac{1}{\sqrt{2\omega}}
       \frac{3 (m_\Delta+m_N) }{ 4m_N( E_N(\bm{q})+m_N)}T_3
\nonumber \\
&\times&
{}  [i G_M(q^2)\bm{S}\times\bm{q}\cdot \bm{\epsilon}
      +G_E(q^2)
       (\bm{S}\cdot\bm{\epsilon} \bm{\sigma}\cdot\bm{q}
   +\bm{S}\cdot\bm{q}\bm{\sigma}\cdot\bm{\epsilon}) \nonumber \\
 & & +
        \frac{G_C(q^2)}{m_\Delta}
      \bm{S}\cdot\bm{q} \bm{\sigma}\cdot\bm{q} \epsilon_0], \label{eqh50}
\end{eqnarray}
where $e=\sqrt{4\pi/137}$, $q=(\omega,\bm{q})$ is the photon four-momentum,
and $\epsilon=(\epsilon_0,\bm{\epsilon})$ is the photon polarization
vector.
The above definition allows us to calculate the multipole amplitudes of
the $\gamma N \rightarrow \Delta$ in terms of $G_M(q^2)$,
$G_E(q^2)$ and $G_C(q^2)$.
Explicitly, we have\cite{sl2}
\begin{eqnarray}
A_M(q^2) & =& [\Gamma_{\gamma N \rightarrow \Delta}]_{M1}=
 N G_M(q^2) \label{eqh51}\\
A_E & = & [\Gamma_{\gamma N \rightarrow \Delta}]_{E2}=
  -NG_E(q^2) \label{eqh52}\\
A_C & = & [\Gamma_{\gamma N \rightarrow \Delta}]_{C2}=
N\frac{\mid \bm{q}\mid}{2 m_\Delta}G_C(q^2), \label{eqh53}
\end{eqnarray}
with 
\begin{eqnarray}
N=\frac{e}{2m_N}\sqrt{\frac{m_\Delta \mid \bm{q}\mid}{m_N}}
\frac{1}{[1-q^2/(m_N+m_\Delta)^2]^{1/2}} \label{eqh54}
\end{eqnarray}
                                                                                
We first discuss the results at $q^2=0$ photon point.
The values of $A_M$, $A_E$, and
$A_C$ determined in Refs.\cite{sl1,sl2}
are listed in Table III. These are the quantities we would like
to interpret within the considered constituent quark model. Assuming that
$\Gamma_{\gamma N,\Delta}$ of Eq. (\ref{eqh30}) is what has been determined in the SL 
model, the values listed in Table III thus include the contribution
not only from the quark-excitation term $\Gamma^0_{\gamma N,\Delta}$ of
 Eq. (\ref{eqh31}), which
can be calculated from Eqs. (\ref{eqh45}) and (\ref{eqh49}), but also include
the pion-loop contributions illustrated in Fig. \ref{fig7}. These loop contributions
can be calculated by inserting appropriate intermediate states in the
commutators of Eqs. (\ref{eqh24})-(\ref{eqh25}). As an example, we write down the 
expression for the mechanism Fig. 7b in the rest frame of $\Delta$
($\bm{p}_\Delta=0$, $\bm{p}_N = -\bm{q}$) 
\begin{eqnarray}
<\Delta|- iS_1 J^\mu_{B',B} iS_1|N>
& & =
 \int d\bm{k} \sum_{B=N,\Delta} \nonumber \\
 & &
\frac{
<\Delta|\Gamma^{0 \dagger}_{\pi\Delta,\Delta}|
                \pi(\bm{k}),\Delta>
<\Delta|J^\mu_{\Delta,B}| B>
<\pi(\bm{k}),B|\Gamma^{0}_{\pi B,N}|N>}
{(m_\Delta -E_\Delta(\bm{k})-E_\pi(\bm{k}))
 (E_N(\bm{q}) - E_B(\bm{q}+\bm{k})-E_\pi(\bm{k}))} \nonumber \\
& & =
 \int d\bm{k}  \sum_{B=N,\Delta} \nonumber \\
 & &
\frac{F_{\pi\Delta,\Delta}^\dagger(0,-\bm{k},\bm{k})
<\Delta|J^\mu_{\Delta,B}| B>F_{\pi B,N}(-\bm{k}-\bm{q},-\bm{q},\bm{k})}
{(m_\Delta -E_\Delta(\bm{k})-E_\pi(\bm{k}))
 (E_N(\bm{q}) - E_B(\bm{q}+\bm{k})-E_\pi(\bm{k}))} \label{eqh55}
\end{eqnarray}
An important point to note here is that the integrand in the 
loop-integration is independent of the collision energy $W$ 
and has no singularity in the integration region $0 \le k \le \infty$.
Thus the included pion cloud effects are different from what
were calculated  from the SL model :
$[v_{\gamma \pi}G_{\pi N}(W)\bar{\Gamma}_{\pi N, \Delta}(W)]$ which depends
on the collision energy $W$ in the $\pi N$ propagator $G_{\pi N}(W)$.
 Qualitatively speaking, the one-loop
contributions of Fig. \ref{fig7} are due to virtual pions which are part of the
internal structure of N or $\Delta$, while the SL model only
accounts for the effects due to pions in scattering states which can reach
the on-shell momentum asymptotically. 

The $Q^2=0$ ($Q^2= - q^2 >0)$
 results from our complete calculations for all one-loop
terms in Fig. \ref{fig7} 
 are presented in Table IV. In the first row,
we list the values from
 $\Gamma_{\gamma N \rightarrow \Delta}^0 $, which
is due to photon interactions with constituent quarks.
As expected, the assumed spherical s-wave
quark configurations do not have $E2$ and $C2$ transitions.
In the same table we also list the contribution from each loop
contribution illustrated in Fig. \ref{fig7}. The terms under 'pion' and
'Seagull' are from Fig. 7a and 7c respectively. Fig. 7d gives the
contribution 'Normalization' which is the consequence of the appearance
of the mass shifts $\Sigma^0_N$ and $\Sigma^0_\Delta$ in the
effective Hamiltonian Eq. (\ref{eqh30}) and is naturally derived here by using the 
unitary transformation method. Fig. 7b contains the contributions due to
spin transition and convection current. Their separate contributions are listed
under 'Spin' and 'Convection' respectively.
We note that 
$A_C$ only has contribution from pion term because of the
angular moment selection rule. 

We should emphasize here that the
present calculations are based on a non-relativistic quark model and can only
be compared qualitatively  with the empirical values (table III)
determined in the SL model.
Thus we should not worry about their differences in absolute magnitudes.
Rather, we focus on the relative importance between $A_M$, $A_E$ and $A_C$ 
listed in Tables III and IV.
 
The first interesting result in Table IV is that
the  total one-loop correction (Total - $\Gamma^0_{\gamma N,\Delta}$) for the
magnetic form factor $G_M$ is only about 4 $\%$ of the 'bare' value
$\Gamma^0_{\gamma N,\Delta}$, 
mainly due to the large
cancellations between different contributions. In particular,
the very large contribution
from 'Normalization' of Fig. 7d plays a crucial role. We also see that
 the calculated 
$A_E$ and $A_C$ in Table IV
are in opposite signs of the values listed in Table III.
These results have the following implications.
 First, the $\Gamma^0_{\gamma N,\Delta}$ values of $A_E$ and $A_M$ in Table IV 
could be nonzero and negative such that the total values become the SL values
listed in Table III. This can be the case 
if we assume that the quark wavefunctions of $\Delta$
and/or $N$ could have a $L=2$ d-state component.
The other possibilities are that there could have multi-pion loop corrections
and exchange current contribution of the quark electromagnetic 
current\cite{buchmann}.
Within our formulation, some of these mechanisms
can be derived from applying
the third-order unitary transformation $U_3$.  
A more detailed study along this line is clearly needed to
make progress.

\begin{table}
\caption[]{'bare' helicity amplitudes of the SL model[1,2].
 Unit is  $10^{-3}GeV^{-1/2}$. }
\begin{tabular}{crrrrrrr}
        \\ \hline
$A_M$ & 173.3  \\
$A_E$ & -2.3 \\
$A_C$ &  -2.2
\end{tabular}
\end{table}
                                                                                
\begin{table}
\caption[]{'bare' helicity amplitudes in quark model with loop
 correction. Unit is  $10^{-3}GeV^{-1/2}$. SL is the result from
Ref.[1,2] }
\begin{tabular}{rrrrrrrrrrrrrr}
      & $\Gamma^0_{\gamma N,\Delta}$ & & Pion && Spin & & Convection && Seagull 
&& Normalization & & Total  \\ \hline
$A_M$ & 204.9  && 23.7 && 24.2  && 0  && -3.3 && -35.6 && 213.8  \\
$A_E$ &  0     && 2.8  &&  -0.1 && 0.1 && 0.9  && 0     && 3.6  \\
$A_C$ &  0     &&1.1  & &   0   && 0.0 & &  0  && 0     && 1.1
\end{tabular}
\end{table}

\begin{figure}[h]
\centering
\includegraphics[width=8cm]{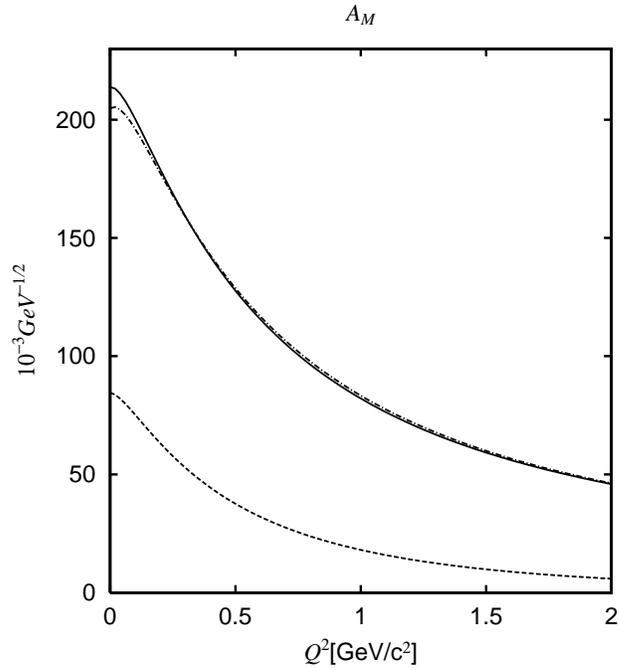}
\caption{ $A_M(Q^2)$. Dot-dashed is from quark excitation 
$\Gamma^0_{\gamma N,\Delta}$, the solid curve is the sum of
$\Gamma^0_{\gamma N,\Delta}$ and $\Gamma^{1-loop}_{\gamma N,\Delta}$.
The dashed curve is from pion scattering calculated in SL model.}
\label{fig8}
\end{figure}

\begin{figure}[h]
\centering
\includegraphics[width=5cm]{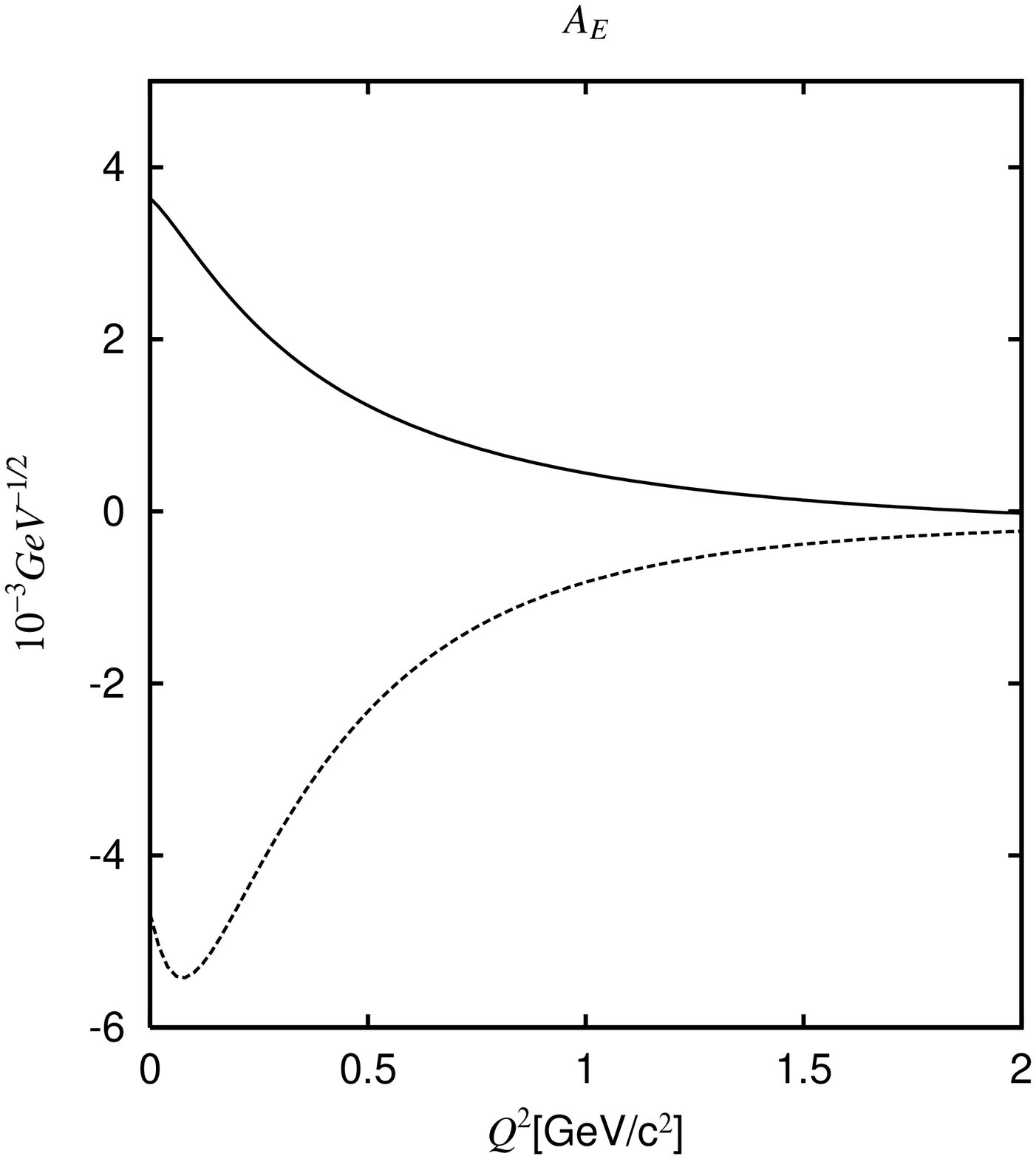}
\includegraphics[width=5cm]{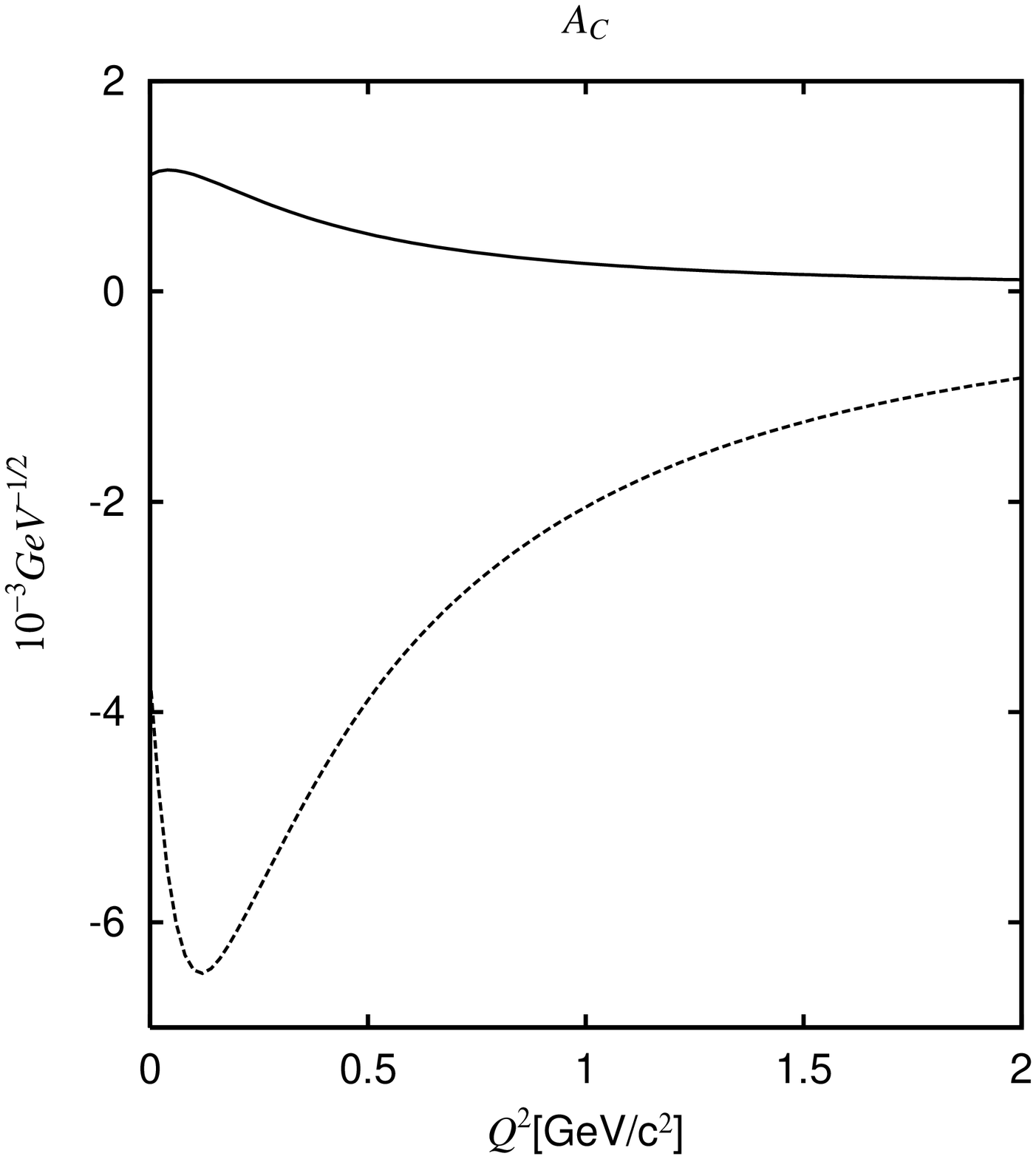}
\caption{ Left:$A_E(Q^2)$, Right:$A_C(Q^2)$. Solid curves are from one-loop
corrections, while the dashed curves are from pion scattering calculated in SL
model.}
\label{fig9}
\end{figure}

The calculated $Q^2-$ dependence of the
$\gamma N \rightarrow \Delta $ form factors are displayed in 
Figs. \ref{fig8} and \ref{fig9}. 
The dominant M1 transition is shown in Fig. \ref{fig8}. The difference 
between the solid and dashed curves is due to the one-loop corrections.
It is very weak and only visible at very low $Q^2$. On the other hand, the
pion cloud effects due to scattering states (dashed curve) is very large,
As discussed in detail in Ref. \cite{sl2}, this finding explains why the 
conventional constituent quark model predictions disagree with the
empirical value of the magnetic M1 transition of $\gamma N \rightarrow \Delta$.
The present result for one-loop corrections do not change that conclusion.

The situation for $A_E(Q^2)$ and $A_C(Q^2)$ is quit different.
Here we do not have contribution from quark excitation 
term $\Gamma^0_{\gamma N,\Delta}$ because of the assumed L=0 wavefunctions.
We see that the calculated
one-loop corrections (solid curves) are comparable in magnitudes
to the pion cloud effects
due to scattering state (dashed curves) calculated in SL model. More
importantly, they have very different $Q^2$-dependence and are
opposite in signs. As seen in Eq. (\ref{eqh30}),
the solid curves must be interpreted as part of the bare form factors
determined phenomenologically in SL model. Namely, the form factors
obtained from subtracting the solid curves from SL model's bare form factors
are the contribution from quark excitation.
This will be an important information for testing various
hadron structure calculations. However, such information can not be
realistically extracted here because of the simplicity of the model
employed.


\section{One-loop corrections on non-resonant $\gamma N \rightarrow \pi N$}

One of the  difficulties the SL model has in describing
the data is from the non-resonant amplitude. 
In this section, we would like to
explore whether this can be improved by including the one-loop corrections
$v^{1-loop}_{\gamma\pi}$ of Eq. (\ref{eqh32}).  As a start, we will
focus on the near threshold region and consider only the
the $E_{0+}$ amplitude. A complete calculation of $v^{1-loop}_{\gamma\pi}$
 for all partial waves
up to $\Delta$ resonance energy
is much more involved and will be explored elsewhere.
 
First, we point out that the SL model failed to describe the
near threshold $\gamma p \rightarrow \pi^0 p$ data.
For example, at $E_\gamma =145$ MeV the SL model gives
(after taking into account the effects due to the mass difference
between $\pi^0$ and $\pi^{\pm}$ )
\begin{eqnarray}
E_{0+}(145 {\it MeV}) 
= -2.47(Born) + 2.31(Rescattering) = -0.15 [10^{-3}/m_{\pi^+}]
\label{eqh56}
\end{eqnarray}
where $Born$ is from the non-resonant production operator
$v^{tree}_{\gamma\pi}$ constructed in SL model, $Rescattering$ include the
effects due to final $\pi N$ interaction.
The empirical value is $E_{0+}^{exp}$ (145 MeV) $\sim $ -1.50.
In getting the above result, we find that the main contributions to the
 Born term are from the nucleon-direct and nucleon-exchange diagrams, while
the rescattering term is mainly from pion-pole and contact interaction
through $\gamma + p \rightarrow \pi^+ + n \rightarrow \pi^0 + p$ charge-exchange
process. We also find that 
the s-wave charge exchange pion rescattering is dominated by the $\rho$-exchange
$\pi-N$ potential and  the Born approximation $t_{\pi N} \sim v_{\pi N}$
is accurate.
Furthermore the short range approximation of $\rho$-exchange potential
($1/(m_\rho^2 + (\bm{p}_N-\bm{p}^\prime_N)^2) \sim 1/m_\rho^2$)
is accurate within 10\% in determining the rescattering effects in the
considered near threshold energy region.
With these considerations,
the one-loop corrections near threshold
can be calculated with the following much simplified Hamiltonian
\begin{eqnarray}
H_I & = & \frac{f_{\pi NN}}{m_\pi}\bar{\psi}_N\gamma_5\gamma_\mu
\partial^\mu \vec{\phi}_\pi\cdot\vec{\tau}\psi_N +
\lambda \bar{\psi}_N\gamma^\mu \vec{\tau}\psi_N \cdot \vec{\phi}_\pi\times
 \partial_\mu \vec{\phi}_\pi \,.
\label{eqh57}
\end{eqnarray}
Here  the second term is a contact interaction with the
strength determined from the $\rho$-exchange coupling constants :
$\lambda=g_{\rho\pi\pi}g_{\rho NN}/(2m_\rho^2)$.
By minimum substitution, the second term of Eq. (\ref{eqh57}) will generate
a interaction current
\begin{eqnarray}
j^{\mu}_{N,N\pi\pi} & = & e\lambda [\bar{\psi}_N\gamma^\mu \vec{\tau}\psi_N
         \times \vec{\phi}_\pi ]\times \vec{\phi}_\pi \,.
\label{eqh58}
\end{eqnarray}
It induces an electromagnetic contact interaction involving two pions.
To maintain the gauge invariance within the model defined by
 the simplified interaction Hamiltonian 
Eq. (\ref{eqh57}), this current is included in the calculation along with the
currents $J_\pi$, $J_{B^\prime, B}$ and $J_{B^\prime, B\pi}$
given in Eqs. (\ref{eqh46})-(\ref{eqh48}) and illustrated in Fig. \ref{fig2}.
All coupling constants and vertex form factors are taken from SL model.

With the above simplified model, we first re-calculate the rescattering 
contributions, $ \sim v^{tree}_{\gamma,\pi}G_{\pi N}(W) v_{\pi N}$ to
the $E_{0^+}$ amplitude
for $\gamma p \rightarrow \pi^0 p$.
The results 
at $E_\gamma =145$ MeV are listed in Table V.
It is instructive to note here that the calculated rescattering contribution
  involves  cancellation between the terms (d) and (e).
The total rescattering value 2.36 is very close to
the value 2.31 of the rescattering term in Eq. (\ref{eqh56}) of the SL model.
 This justifies
the use of the simplified model defined by Eqs. (\ref{eqh57}) and (\ref{eqh58}).

\begin{table}[htb]
\caption[]{ Rescattering contributions to the
$E_{0^+}$ amplitude of $\gamma p \rightarrow \pi^0 p$
at 145 MeV, calculated from mechanisms (b)-(e)
illustrated in Fig. \ref{fig6} using the
model defined by Eqs. (\ref{eqh57})-(\ref{eqh58}).}
\begin{tabular}{cccccccccccccccc}
Diagram & & &  (b) & & & (c) & & &(d) & & &(e)  & & & sum \\ \hline
 & & &-0.074  & & &-0.685  & & & -1.966 & & & 5.087 & & &2.36  \\ \hline
\end{tabular}
\end{table}

\begin{figure}[h]
\centering
\includegraphics[width=10cm]{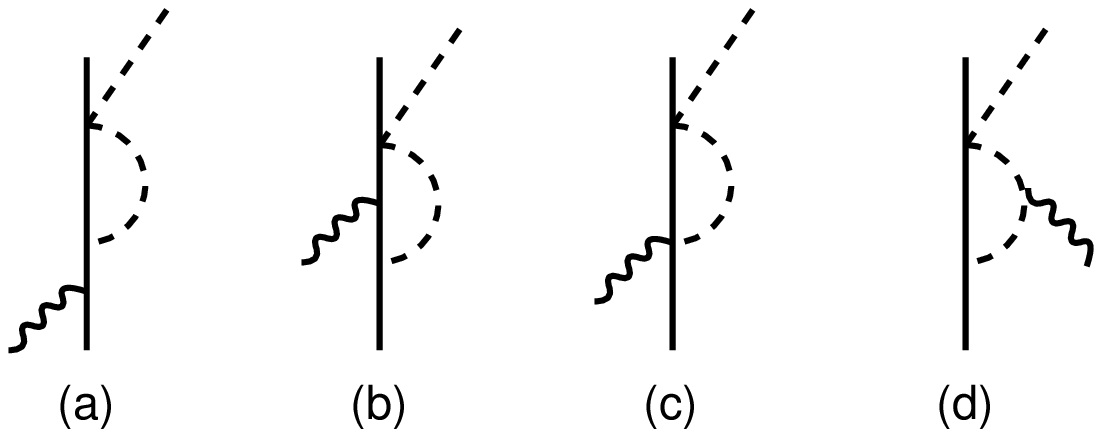}
\caption{Subset of loop corrections on the $\gamma N \rightarrow \pi N$
transition amplitude.  } 
\label{fig10}
\end{figure}

\begin{figure}[h]
\centering
\includegraphics[width=10cm]{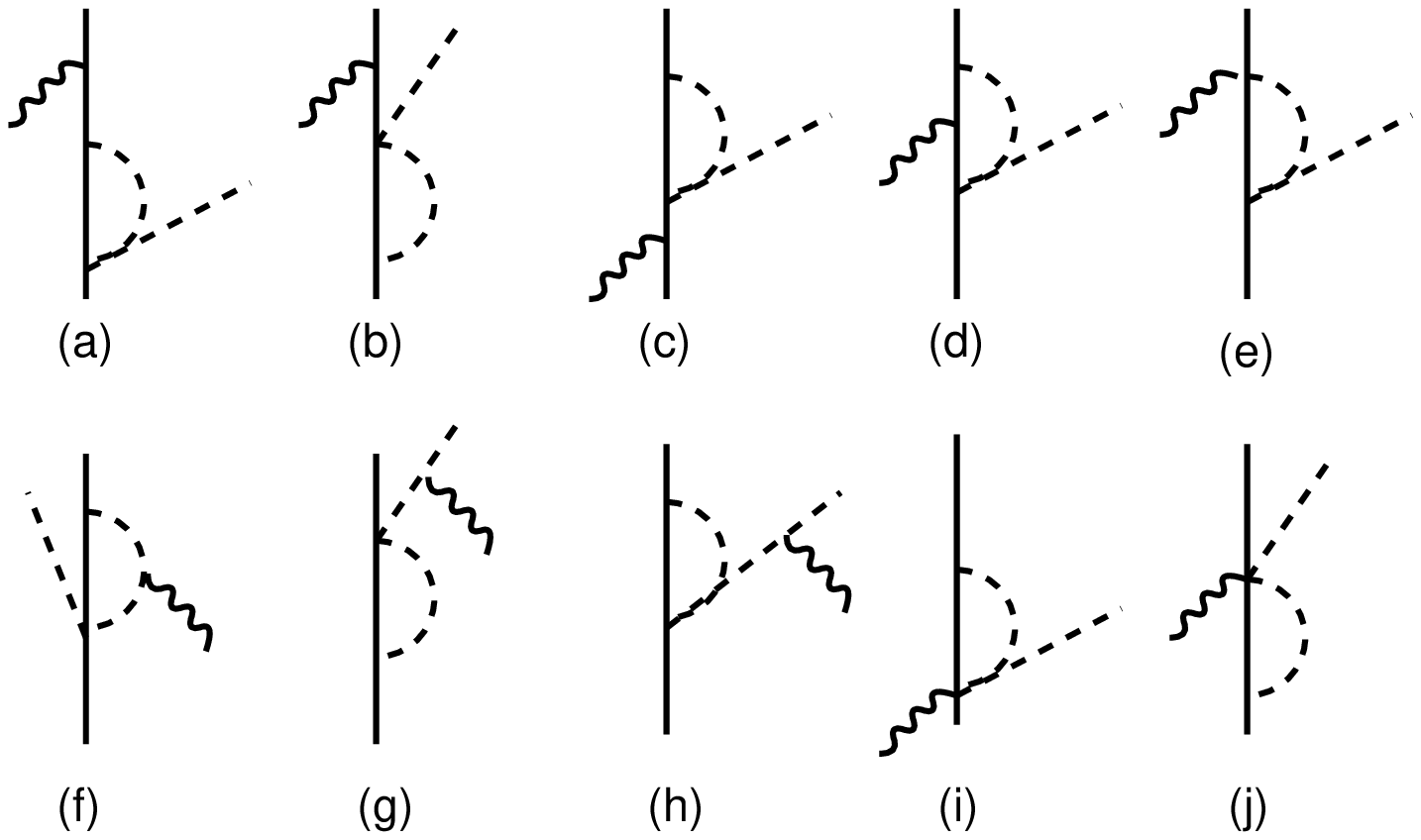}
 \caption{Subset of
Loop corrections on the $\gamma N \rightarrow \pi N$ transition amplitude. }
\label{fig11}
\end{figure}

The one-loop corrections can be calculated from Eq. (\ref{eqh23}) by
 inserting appropriate intermediate states. The resulting
amplitudes are illustrated in Figs. \ref{fig10} and \ref{fig11}.
Note that diagrams
in Figs. \ref{fig10} and \ref{fig11} are not time-ordered diagrams. 
Rather they just illustrate
the structure of the matrix element of each term in $v^{1-loop}_{\gamma\pi}$. 
As seen from Eq. (\ref{eqh14}) and Eq. (\ref{eqh17}), 
the loop integrations for all processes in Figs. \ref{fig10} and \ref{fig11} will involve
$energy-independent$ propagators associated with these two operators.
Thus, although the  diagrams in Fig. \ref{fig10} look similar to the
rescattering terms, but they are $energy-independent$ matrix elements.
The calculations for these loops are tedious but straightforward,
and will not be elaborated here.

\begin{table}[htb]
\caption[]{ one-loop contributions to the
$E_{0^+}$ amplitude of $\gamma p \rightarrow \pi^0 p$
at 145 MeV, calculated from mechanisms illustrated in Fig. 11 using the
model defined by Eqs.(54).}
\begin{tabular}{ccccccccccccccccc}
 & Diagram & & &  $E_{0+}$ & & & Diagram  & & & $E_{0+}$ & & &  & & & \\ \hline
Fig. 10 & (a) & & &  -0.079  & & &  (c)  & & & -0.024  & & & & & & \\
        & (b) & & &  -0.090  & & &  (d)  & & &  0.475  & & & & & & \\ \hline
Fig. 11 & (a) & & &  0.157  & & &  (f)  & & & -0.418  & & & & & & \\
        & (b) & & & -1.192  & & &  (g)  & & &  0.00   & & & & & & \\
        & (c) & & &  0.875  & & &  (h)  & & &  0.00   & & & & & & \\
        & (d) & & & -0.085  & & &  (i)  & & & -0.696  & & & & & & \\
        & (e) & & & -1.011  & & &  (j)  & & &  0.699  & & & & & & \\  \\ \hline
        &   & & &    & & & Sum = -1.39 & & & & & & & & &
\end{tabular}
\end{table}

\begin{figure}[h]
\centering
\includegraphics[width=8cm]{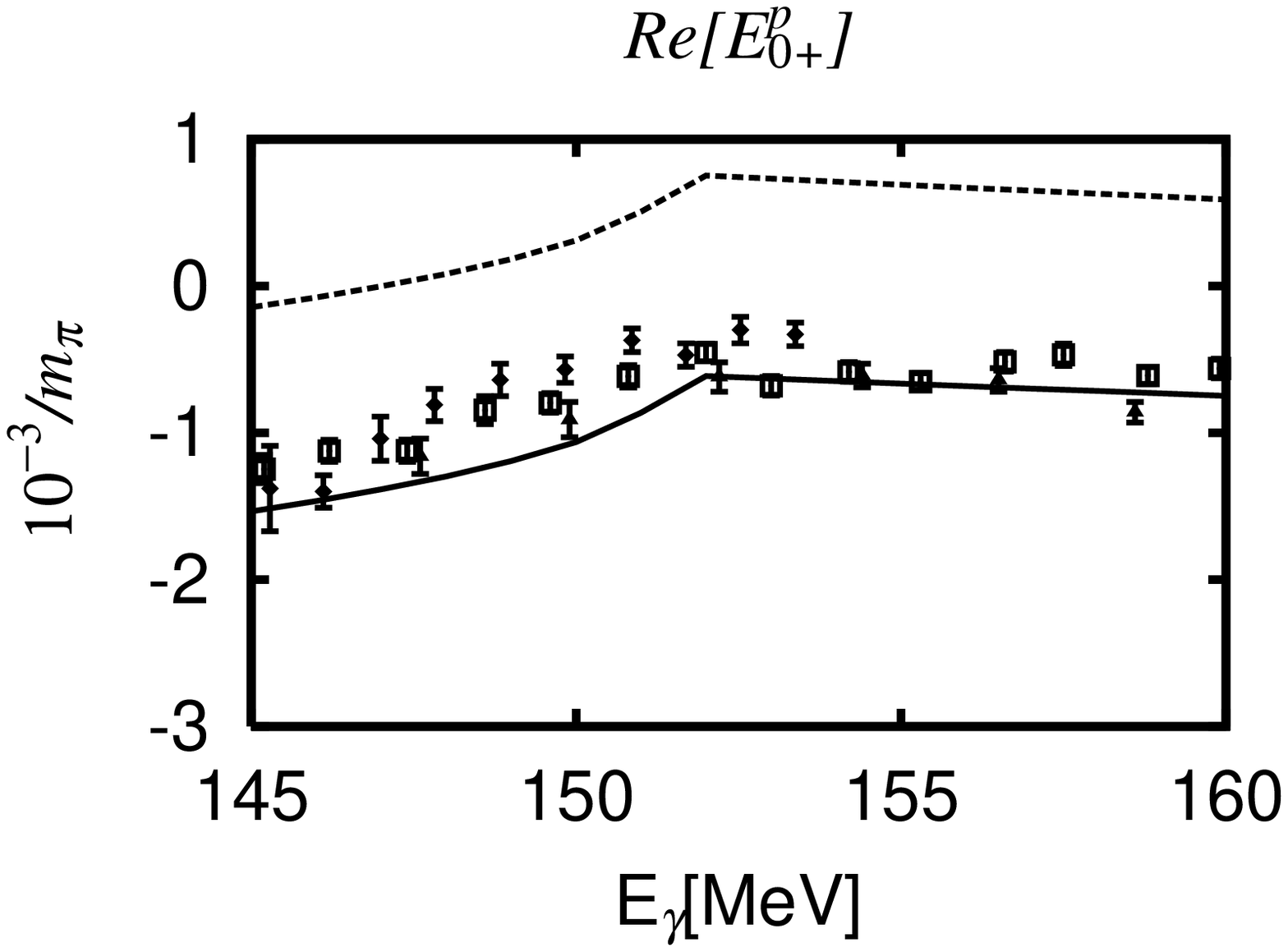}
\includegraphics[width=8cm]{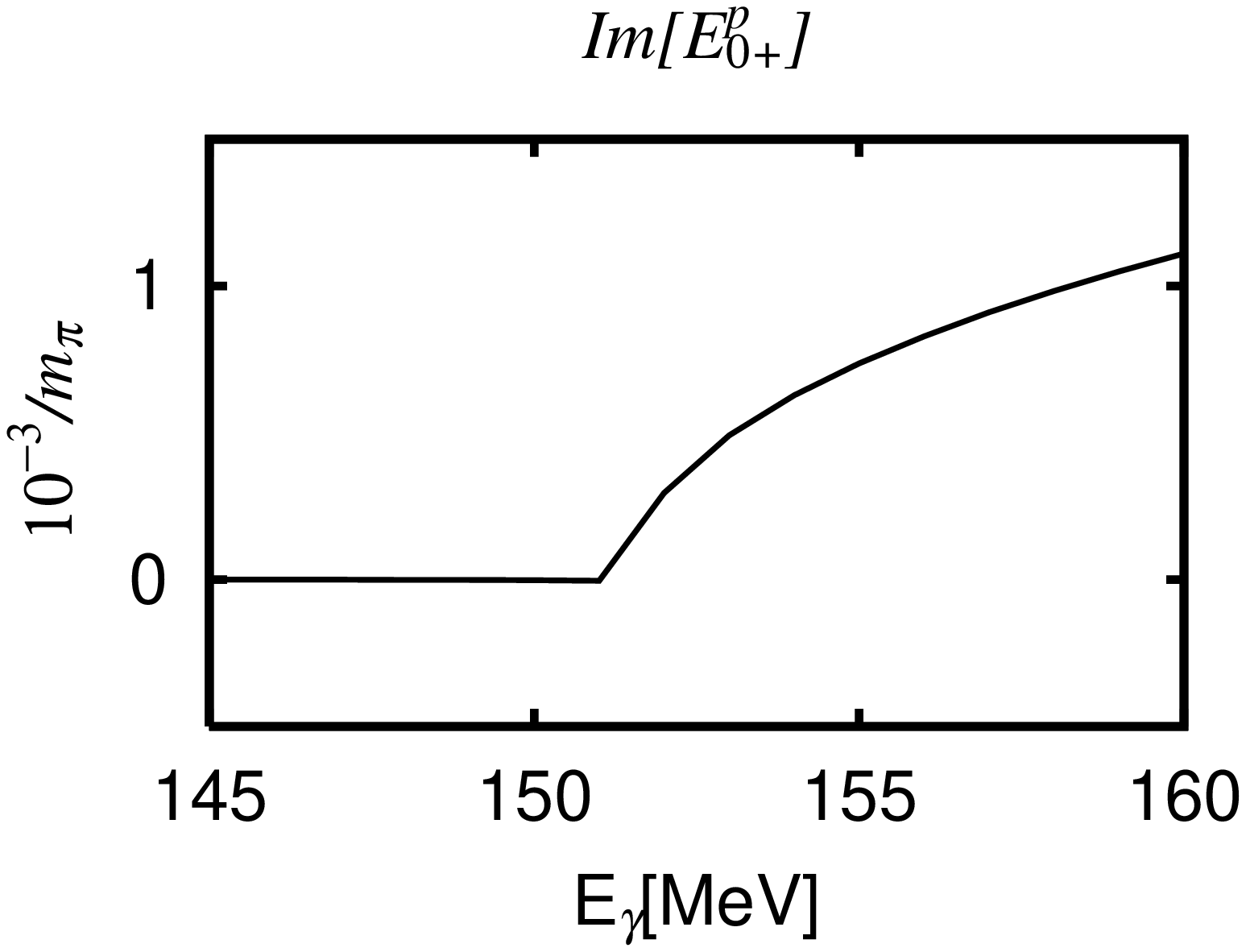}
 \caption{
 $E_{0+}$ amplitude of $\gamma p \rightarrow \pi^0 p$. The dotted curve is
from $v^{tree}_{\gamma \pi}$. The solid curves are obtained when
the one-loop corrections $v^{1-loop}_{\gamma\pi}$ are included. The
data are from Refs \cite{e0exp1,e0exp2,e0exp3}}
\label{fig12}
\end{figure}

Our  results  at $E_\gamma =145$ MeV
for each of the one-loop corrections shown in
 Figs. 10 and 11 are listed in Table VI. 
The results listed in Tables V and VI lead to
\begin{eqnarray}
 E_{0+}(145 MeV)= -2.47(Born) + 2.32(Rescattering) - 1.39(Loop) = -1.54
\label{eqh59}
\end{eqnarray}
This reproduces the empirical value $E_{0+}^{exp}$ (145 MeV) $ \sim$ -1.50.
The calculated effect of the one-loop corrections for
 $E_{0+}$ in the near threshold energy region is shown in
 in Fig. \ref{fig12}. Clearly, the one-loop corrections
 drastically reduce  the magnitudes and
bring the results to agree with the empirical values. The kinks due to
the cups effect are reproduce well in our calculations.

\begin{figure}[h]
\centering
\includegraphics[width=6cm]{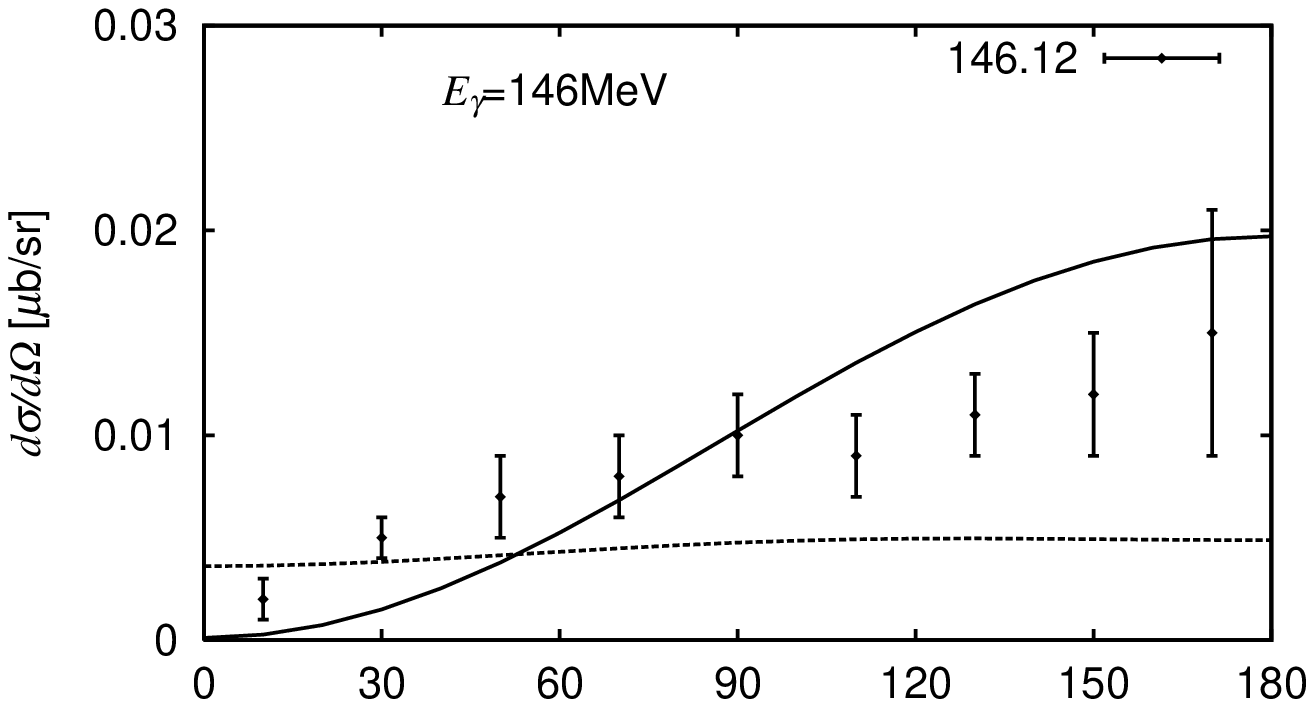} 
\includegraphics[width=6cm]{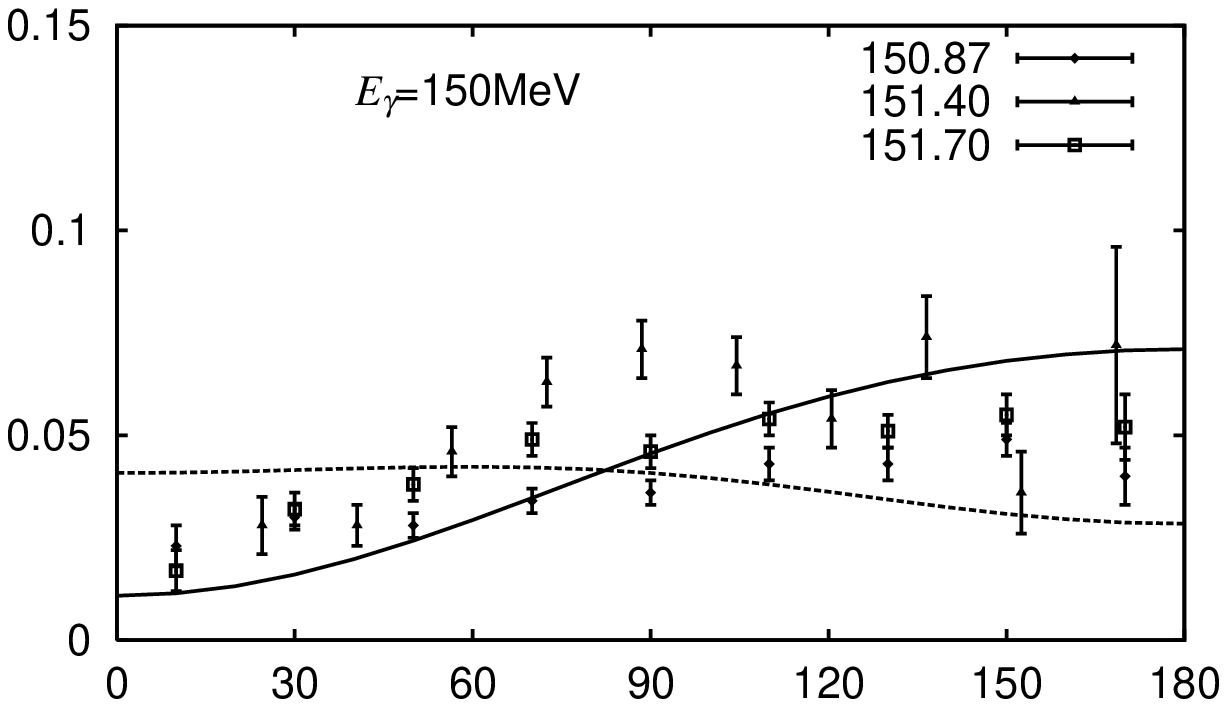} 
\includegraphics[width=6cm]{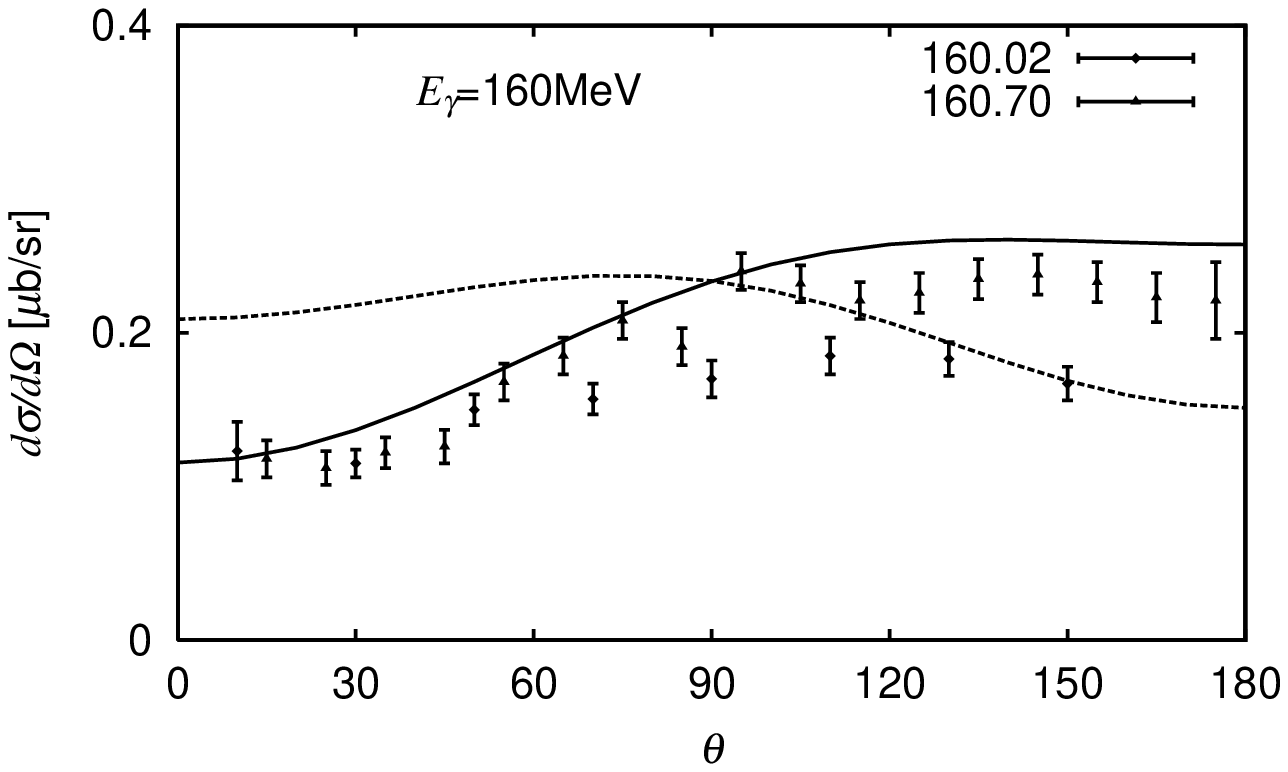}
\includegraphics[width=6cm]{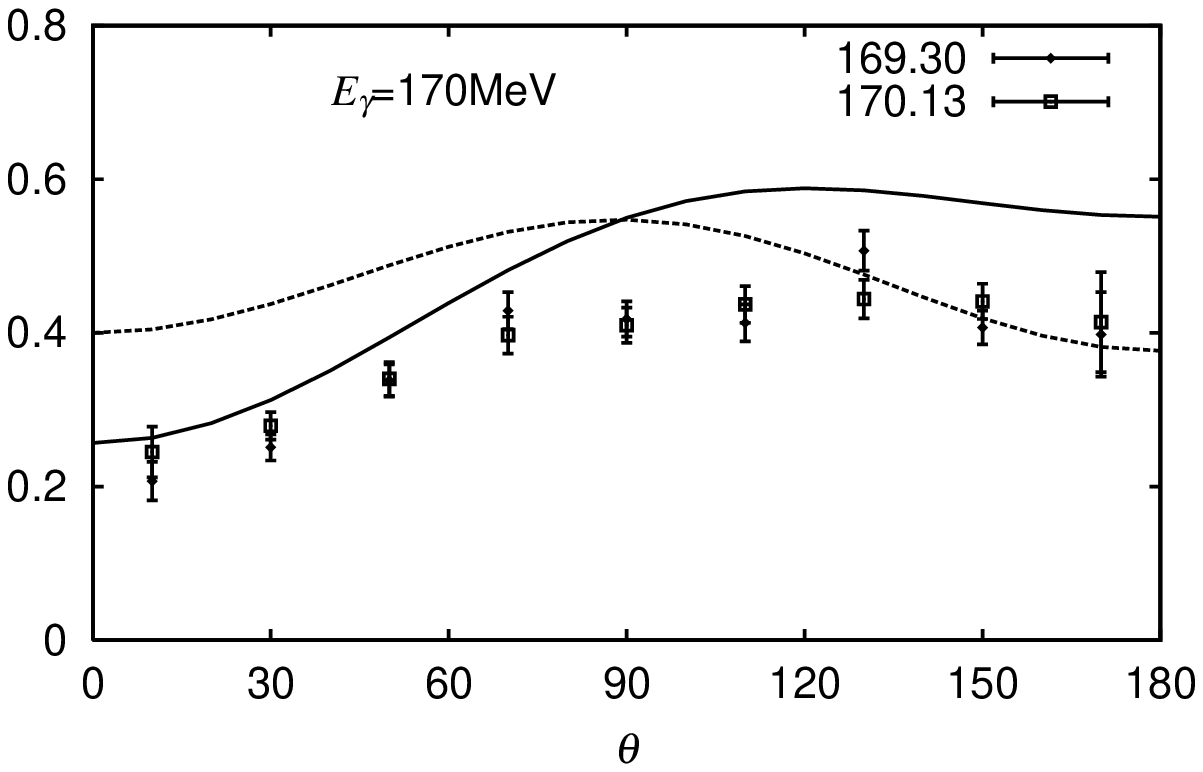}
\caption{$\gamma p \rightarrow \pi^0 p$ at 146, 150, 160, and 170 MeV.
Solid curves are obtained when the one-loop corrections 
$v^{1-loop}_{\gamma\pi}$ are included. Data are from Ref.\cite{e0exp1,e0exp2}.}
\label{fig13}
\end{figure}

In Fig. \ref{fig13}, we show that 
the one-loop corrections on the $E_{0^+}$ amplitude 
can change significantly the calculated 
angular distributions to better agree with the data.
To see the full one-loop correction effects, we need to also calculate
other multipole amplitudes. This along with the results for the $\Delta$
region will be explored elsewhere.

\section{Summary and Outlook}

In this paper, we have applied the unitary transformation to derive 
the leading-order corrections on the effective
Hamiltonian of the SL model for electromagnetic pion production reactions.
We have investigated the
one-loop corrections on the masses of $N$ and $\Delta$,
the $\gamma N \rightarrow \Delta$ vertex, and the non-resonant pion production operators.
Qualitatively speaking, the derived one-loop corrections are due to the
 virtual pions which
are part of the internal structure of $N$ or $\Delta$, while the pion cloud effects
generated within the SL model or the other dynamical
models, such as the Dubna-Mainz-Taipei (DMT) model \cite{dmt},
 only account for the effects 
due to pions in the scattering states which can reach the on-shell
 momentum asymptotically.

With the one-loop corrections included in determining the mass parameters, we find that the
free Hamiltonian of the model can be identified with the conventional constituent quark model.
We then proceed to apply such a constituent quark
model to calculate the one-loop corrections on the $\gamma N \rightarrow \Delta$
transition form factors. It is found that the one-loop corrections on the magnetic M1
transition  is very small. Our results further establish the conclusion
reached by the SL model that the large discrepancy between the conventional constituent quark
model predictions and the empirical values are due to the pion cloud effects 
associated with the pions in scattering states. 

The calculated one-loop contributions to the electric E2 ($A_E$) and Coulomb C2 ($A_C$)  
form factors of the $\gamma N \rightarrow \Delta$ transition are found to be
in opposite signs of that due to pion cloud associated with the scattering 
states. One possible implications of this result is that
 the extracted empirical values of SL model
could be largely due to the
nonspherical $L=2$ intrinsic quark excitations which could lead
to nonzero and negative contributions to  $A_E$ and $A_C$.
On the other hand, there could have
higher-order exchange current contributions which are not included in this
work, but must be also calculated
for a complete understanding of the empirical values of SL model.
Clearly more works are  highly desirable.

We have also found that the one-loop corrections on 
the non-resonant pion production operator can
resolve the difficulty the  SL model encountered in reproducing the empirical
$E_{0+}$ amplitude of near threshold $\pi^0$ photoproduction. It will be worthwhile to
further extend this work to calculate these one-loop corrections 
for higher partial waves.
Some of the discrepancies between the SL model and the data in the $\Delta$ excitation
could be removed by including these corrections. 
Our effort in this direction will be reported
elsewhere.

To end, we emphasize that the SL model is obtained from 
keeping only the lowest order terms of
a formulation within which the higher order terms can be rigorously derived.
 Attempts to fit the data by adjusting the current SL model
are not justified theoretically.
The most important task to improve the SL model is to include these corrections 
order by order until the convergence of  the predictions has achieved.
In this work we have taken a very first step in this direction. 
Undoubtly, much more works
are needed to complete a consistent dynamical model of electromagnetic pion 
production reactions.

\vspace{2cm}

This work was supported by the U.S. Department of Energy, Office of
Nuclear Physics Division, under contract no. W-31-109-ENG-38, 
and by Japan Society for the Promotion
of Science, Grant-in-Aid for Scientific Research (C) 15540275.

\end{document}